\documentclass{elsart}

\usepackage{natbib}
\usepackage{graphics}
\usepackage{graphicx}
\usepackage{epsfig}
\usepackage{amssymb}

\def\Box{\mbox{ }\rule[0pt]{1.5ex}{1.5ex}}

\input{epsf.sty}
\usepackage{latexsym,amsmath,amsfonts,amscd}
\usepackage{changebar}

\begin{document}

\begin{frontmatter}



\title{A WENO Algorithm of the Temperature and Ionization Profiles
     around a Point Source}

\author[label1]{Jing-Mei Qiu},
\author[label2,label3]{Long-Long Feng},
\author[label1]{Chi-Wang Shu},
\author[label4]{Li-Zhi Fang}

\address[label1]{Division of
Applied Mathematics, Brown University, Providence, RI 02912, USA}
\address[label2]{Purple Mountain Observatory, Nanjing, 210008,
P.R. China}
\address[label3]{National
Astronomical Observatories, Chinese Academy of Science, Chao-Yang
District, Beijing 100012, P.R. China}
\address[label4]{Department of Physics, University of Arizona,
Tucson, AZ 85721, USA}


\begin{abstract}

We develop a numerical solver for radiative transfer problems based
on the weighted essentially nonoscillatory (WENO) scheme modified
with anti-diffusive flux corrections, in order to solve the
temperature and ionization profiles around a point source of photons
in the reionization epoch.  Algorithms for such simulation must be able
to handle the following two features: 1. the sharp profiles of
ionization and temperature at the ionizing front (I-front) and the
heating front (T-front), and 2. the fraction of neutral hydrogen
within the ionized sphere is extremely small due to the stiffness
of the rate equations of atom processes. The WENO scheme can properly
handle these two features, as it has been shown to have high order
of accuracy and good convergence in capturing discontinuities and
complicated structures in fluid as well as to be significantly
superior over piecewise smooth solutions containing discontinuities.
With this algorithm, we show the time-dependence of the preheated shell
around a UV photon source. In the first stage the I-front and
T-front are coincident, and propagate with almost the speed of light.
In later stage, when the frequency spectrum of UV photons is hardened,
the speeds of propagation of the ionizing and heating fronts are both
significantly less than the speed of light, and the heating front
is always beyond the ionizing front. In the spherical shell
between the I- and T-fronts, the IGM is heated, while atoms keep
almost neutral. The time scale of the preheated shell evolution
is dependent on the intensity of the photon source. We also find that
the details of the pre-heated shell and the distribution of neutral
hydrogen remained in the ionized sphere are actually sensitive to
the parameters used. The WENO algorithm can provide stable and robust
solutions to study these details.

\end{abstract}


\end{frontmatter}

\section{Introduction}

The profile of HII region around an isolated source of UV
photons is an old topic in astrophysics. A classical result was
given by Str\"omgren (1939), who showed that the profile of the
spherical HII region of a point source embedded in a uniformly
distributed hydrogen gas with number density $n$ at radial
coordinate $r=0$ can be approximately described by a step
function as
\begin{equation}
\label{eq1}
f_{\rm HII}(r)\simeq \theta[R_s-r] = \left \{ \begin{array}{ll}
                            1,& {\rm if \ } r<R_s \\
                            0, & {\rm if \ } r>R_s \\
                           \end{array}
                              \right .
\end{equation}
where the fraction of ionized hydrogen $f_{\rm HII}(r)\equiv n_{\rm
HII}(r)/n$, $n_{\rm HII}(r)$ being the number density of ionized
hydrogen. That is, hydrogen gas is sharply divided into two regions:
within a sphere with Str\"omgren radius $R_s$ around the source,
hydrogen is fully ionized, while outside the sphere hydrogen atoms
remain neutral. $R_s$ is determined by the balance between
ionization and recombination
\begin{equation}
\label{eq2}
R_s=\left (\frac {3 \dot{N}}{4\pi \alpha_{B}n^2}\right)^{1/3},
\end{equation}
where $\dot{N}$ is the emission of ionizing photons of the source, and
$\alpha_{B}$ is the recombination coefficient of HII (Osterbrock
\& Ferland 2005). The sharp
boundary at $r=R_s$ is the ionization front (I-front) separating
the HII and HI regions.

The problem of the Str\"omgren sphere has once again attracted many
studies recently, because the formation of HII regions around high
redshift quasars, galaxies and first stars is crucial to
understanding the evolution of the reionization (Cen \& Haiman 2000;
Madau \& Rees 2000; Ciardi et al. 2001; Ricotti et al. 2002; Wyithe
\& Loeb, 2004; Kitayama, et al. 2004; Whalen et al. 2004; Yu, 2005;
Yu \& Lu 2005; Alvarez et al. 2006, Iliev et al. 2006). Unlike the
static solution eq.(\ref{eq1}), new studies focus on the dynamical
behavior of the ionized sphere, such as the time-dependence of the
ionization profile $f_{\rm HII}(t,r)$, the propagation of the
I-front. Besides the HII region and the I-front, a high kinetic
temperature region also exists around the UV photon source due to
the photon-heating of gas. Similar to the I-front, there is also a
T-front separating heated and un-heated gas. The temperature profile
$T(t,r)$ and the T-front are important for probing reionization. For
instance, the region with high kinetic temperature $T$ and low
$f_{\rm HII}$ would be the 21 cm emission region associated with
sources at the reionization epoch (Tozzi et al. 2000; Wyithe et al.
2005; Cen, 2006; Chen \& Miralda-Escude 2006). Although the possible
existence of a 21 cm emission shell around high redshift quasars and
first stars has been addressed qualitatively or semi-analytically in
these references, a serious calculation seems to be still lacking.

Many numerical solvers for the radiative transfer equation have been
proposed (Ciardi et al. 2001, Gnedin \& Abel 2001, Sokasian et al. 2001,
Nakamoto et al. 2001; Razoumov et al. 2002, Cen 2002, Maselli et al. 2003,
Shapiro et al., 2004; Rijkhorst et al. 2005; Mellema et al. 2006;
Susa 2006, Whalen \& Norman
2006). These solvers provide numerical results of the I-front. However,
the results are still diverse due to the usage of different approximations.
Some of the results show that the time scale of the I-front evolution is sensitively
dependent on the intensity of source (White et al. 2003), while some
yield intensity-independent evolution (e.g. Mellema et al. 2006). This
is because the retardation of photon propagation is ignored in the later,
while the recombination is ignored in the former. Therefore, it is worth
to re-calculate this problem without above-mentioned assumptions. It
has been pointed out that to
study the dynamical features of the I- and T-fronts, one would need to
apply high-resolution shock-capturing schemes similar to those developed
in fluid dynamics (Razoumov \& Scott 1999). The finite difference WENO
scheme is an algorithm satisfying this requirement.

Moreover, although the fraction of the remained neutral hydrogen within
the ionized sphere is extremely small, it is not zero. The small
fraction is important to estimate the Ly-$\alpha$ photon leaking at high
redshifts. Therefore, a proper algorithm for the dynamical
properties of the ionized sphere should be able to, on the one hand,
effectively capture sharp profile of ionization and temperature around
the I- and T-fronts, and, on the other hand, give a precise value of
the remained neutral hydrogen between the discontinuities. This can
also be satisfied by the WENO algorithm.

The WENO algorithm has proved to have high order of accuracy and
good convergence in capturing discontinuities and complicated structures
in fluid as well as to be significantly superior over piecewise smooth
solutions containing discontinuities (Shu 2003). We have shown that the
WENO algorithm is indeed effective for solving radiative transfer
problem with discontinuities with high accuracy. For instance, it can
follow the propagation of a sharp I-front and the step function
cut-off of retardation (Qiu et al. 2006). We now develop this method to
solve both ionization and temperature profiles. It is not a trivial
generalization. Because the rate equations of heating-cooling and
ionization-recombination are stiff, the time integration of the WENO scheme
cannot be directly implemented. A
proper  strategy to save
computational cost on time integration will be developed. A
long-term motivation of this work, as we mentioned in (Qiu et al.
2006), is to develop a WENO solver of hydrodynamic/radiative transfer
problems, similar to the development of a
hybrid algorithm of hydrodynamic/N-body simulation based on WENO scheme
(Feng et al. 2004).

The paper is organized as follows. Section 2 describes the problem and
equations needed to be solved. Section 3 presents the WENO numerical
scheme. Section 4 gives the solutions of the temperature and
ionization profiles and the evolution of energy spectrum of photons.
A discussion and conclusion
are given in Section 5.  Details of the atomic processes are listed in
the Appendix.

\section{Basic Equations}

To demonstrate the algorithm, we consider the ionization of a
uniformly distributed hydrogen gas in space ${\bf x}$ with number
density $n$ by a point UV photon source located at the center $|{\bf
x}|=r=0$. Adding helium component in the gas is straightforward and
will not change the algorithm for radiative transfer. If the time
scale $t$ of the growth of the ionized sphere is less than $1/H(t)$,
$H(t)$ being the Hubble parameter, the expansion of the universe can
be ignored. The radiative transfer equation of the specific
intensity $J(t,r,\nu)$ is then (see the Appendix)
\begin{equation}
\label{eq3}
        {\partial J\over\partial (ct)} + \frac {1}{r^2}
        \frac{\partial}{\partial r}\left (r^2J \right )
          = - k_{\nu} J + S
\end{equation}
where $\nu$ is the frequency of photon. The source term, $S$,
is given by
\begin{equation}
\label{eq4}
S(t,|{\bf x}|,\nu)=\dot{E}(\nu)\delta({\bf x})
\end{equation}
where $\dot{E}(\nu)d\nu$ is the energy distribution of photons
emitted by the central source per unit time within the frequency
range from $\nu$ to $\nu+d\nu$. We assume the energy spectrum of UV
photons to be of a power law
$\dot{E}(\nu)=\dot{E_0}(\nu_0/\nu)^{\alpha}$, and $\nu_0$ is the
ionization energy of the ground state of hydrogen $h\nu_0=13.6$ eV.
Integration of $\dot{E}$ over $\nu$ gives the total intensity
(energy per unit time) of ionizing photons emitted by the source,
$\dot{E}=\int_{\nu_0}^{\infty} \dot{E}(\nu)d\nu = \dot{E}_0
\nu_0/(\alpha-1)$.

The absorption coefficient of eq.(\ref{eq3}) is $k_{\nu} =
\sigma(\nu)n_{{\rm HI}}(t, {\bf x})$, where the cross section
$\sigma(\nu)\simeq\sigma_0(\nu_0/\nu )^3$ and $\sigma_0=6.3\times
10^{-18}$  cm$^2$. The evolution of the number density of neutral
hydrogen HI, $n_{\rm HI}(t, r)$, is governed by the ionization
equation,
\begin{equation}
\label{eq5}
\frac{df_{\rm HI}}{dt}=\alpha_{\rm HII}n_ef_{\rm HII}
-  \Gamma_{\rm \gamma HI}f_{\rm HI}- \Gamma_{\rm e HI} n_ef_{\rm HI}
\end{equation}
where $f_{\rm HI}\equiv n_{\rm HI}(t, r)/n$ is the fraction of
neutral hydrogen, and $n_e$ is the number density of electrons.
Obviously, $f_{\rm HI}(t,r)+f_{\rm HII}(t,r)=1$. In eq.(\ref{eq5}),
$\alpha_{\rm HII}$ is the recombination coefficient and $\Gamma_{\rm
eHI}$ is the collision ionization rate. The photoionization rate
$\Gamma_{\rm \gamma HI}(t, r)$ is given by
\begin{equation}
\label{eq6}
 \Gamma_{\rm \gamma HI}(t, r)=
\int_{\nu_0}^{\infty} d\nu \frac{J(t,r, \nu)}{h\nu}\sigma(\nu).
\end{equation}
The kinetic temperature of baryon gas is determined by the equation
\begin{equation}
\label{eq7}
n k_B\frac{dT}{dt} =H-n^2C
\end{equation}
where $k_B=1.38\times 10^{-16}$ erg $K^{-1}$ is the Boltzmann constant and
the temperature $T$ is in unit of K.
The details of the heating $H$ and cooling $C$ are
given in the Appendix.

\section{Numerical Algorithm}

\subsection{Dimensionless variables}

In the numerical implementation, it is convenient to introduce the
dimensionless variables of time, space and frequency defined by
$t'=c\sigma_0 nt$, $r'=\sigma_0 nr$ and $\nu' = \nu / \nu_0$.
$1/\sigma_0 n$ is the optical depth of ionizing photon in neutral
hydrogen gas with density $n$. Therefore, $t'$ and $r'$ are
respectively, the time and distance in units of mean free flight
time and mean free path of ionizing photon $h\nu_0$ in the
non-ionized background hydrogen gas $n$. For the $\Lambda$CDM model,
$n=1.88\times 10^{-7}(1+z)^3$ cm$^{-3}$, where $z$ is redshift,
$t'=0.89 (1+z)^{-3} t$ Myrs and $r'=0.27(1+z)^{-3}r$ Mpc.
Correspondingly, the intensity is rescaled by $J
d\nu=(1/r^2)(h\nu_0/\sigma_0^3 n)J'd\nu'$. Thus, eqs.(\ref{eq3}),
(\ref{eq5}), (\ref{eq6}) and (\ref{eq7}) become
\begin{equation}
\label{eq8}
{{\partial J'}\over{\partial t'}}+{{\partial
J'}\over{\partial r'}}
   =-\left(\frac{1}{\nu'}\right )^3 {f_{\rm HI}} J',
\end{equation}
and
\begin{equation}
\label{eq9}
 c\sigma_0\frac{df_{\rm HI}}{dt'}=\alpha_{\rm
HII}f^2_{\rm HII} -
  \frac{\Gamma_{\rm \gamma HI}}{n}f_{\rm HI}- \Gamma_{\rm e HI}
    (1-f_{\rm HI})f_{\rm HI}.
\end{equation}
\begin{equation}
\label{eq10}
 \frac{\Gamma_{\rm \gamma HI}(t, r)}{n}=
\frac{1}{r'^2} \int_{1}^{\infty} d\nu' \frac{J'(t,r,
\nu')}{\nu'}\frac{1}{\nu'^3}
\end{equation}
\begin{equation}
\label{eq11}
 c\sigma_0k_B\frac {dT}{dt'} =\frac{1}{n^2}H-C.
\end{equation}
where we have assumed $n_e/n=f_{\rm HII}$ and
\begin{equation}
\label{eq1112}
\frac{1}{n^2} H = \frac{h \nu_0}{r'^2}f_{HI} \int_1^\infty \frac{\nu'-1}{\nu'^4}J' d\nu'
\end{equation}

The point source condition eq.(\ref{eq4}) requires $\dot{E}(\nu)d\nu=4\pi
(h\nu_0/n\sigma_0^3)J'(t,0,\nu')d\nu'$. If we take $J'(t,0,\nu') =
J_0(1/\nu')^{\alpha}$, the total intensity of the source is given
by
\begin{equation}
\label{eq12}
\dot{E}= \frac{ 4\pi h\nu_0}{(\alpha-1)n\sigma_0^3}J_0=5.8\times
   10^{42}\frac{1}{\alpha-1}\left (\frac{10}{1+z_r}\right )^3
\left (\frac{J_0}{10^{-3}}\right ) \hspace{5mm} {\rm erg/s}.
\end{equation}

In our numerical calculation, we solve the system of equations
(\ref{eq8}), (\ref{eq9}) and (\ref{eq11}) for the specific intensity
$J'$, the fraction of the neutral hydrogen $f_{\rm HI}$ and the
temperature
$T$ as functions of the radius $r'$, frequency $\nu'$ and time $t'$.
We will drop the prime in the variables $J'$, $r'$, $\nu'$ and $t'$
in this section below, when there is no ambiguity,
and keep prime in the variables in showing the numerical results.

To solve the radiative transfer equation, we adopt the fifth-order
finite difference WENO scheme with anti-diffusive flux corrections.
The fifth-order finite difference WENO scheme was
designed in (Jiang $\&$ Shu 1996) and the anti-diffusive flux
corrections to the high order finite difference scheme was designed
in (Xu $\&$ Shu 2005). The objective of the anti-diffusive flux
corrections is to sharpen the contact discontinuities in the
numerical solution of the WENO scheme as well as to maintain high
order accuracy. A fourth order quadrature formula is used in the
computation of integration in equations (\ref{eq10}) and
(\ref{eq1112}). The third order TVD Runge-Kutta time discretization is
used in time integration for the system of equations (\ref{eq8}),
(\ref{eq9}) and (\ref{eq11}). We now describe our numerical
algorithm in more detail.

\subsection{The computational domain}

The computational domain is $(r, \nu) \in [0, r_{max}]
\times [1, \nu_{max}]$,
where $r_{max}$ and $\nu_{max}$ are chosen such that
$J(r, \nu, t)\approx 0$
for
$r >  r_{max}$ or $\nu > \nu_{max}$. In our computation,
$r_{max} = 1200$ and
$\nu_{max}=10^6$. The computational domain is discretized into a
uniform mesh
in the $r$-direction and into a smooth non-uniform mesh in the
$\nu$-direction.
The uniform mesh in the $r$-direction is
$$
r_i = i \Delta r \quad with \quad \Delta r = r_{max}/N_r,
\quad i=0, ..., N_r
$$
and the non-uniform mesh in the $\nu$-direction is taken to be
$$
\nu_j = 2^{\xi_j}\quad with \quad \xi_j=j \Delta \xi, \quad
\Delta \xi = {\log_2}\nu_{max}/N_\nu,\quad
j=0, ... , N_\nu
$$
which is allowed because only integration, e.g. in equations (\ref{eq10})
and (\ref{eq1112}),
is involved in the computation with respect to the $\nu$-variable. In
our model, large $\nu$
contributes little in equations (\ref{eq10}) and (\ref{eq1112}), therefore
the non-uniform
mesh is designed
in a way such that the mesh becomes coarser for larger $\nu$.

\subsection{Approximation to the spatial derivative}

To approximate the spatial derivative in equation (\ref{eq8}),
the WENO scheme with anti-diffusive flux corrections is used.
Specifically, to calculate $\partial J/\partial r$, the variable $\nu$
is fixed and the approximation is performed along the $r$-line
\begin{equation}
\label{shuadd60}
{\partial \over {\partial r}}J(t^n, r_i, \nu_j) \approx
\frac{1}{\Delta r} \left( {\hat{h}^a}_{i+1/2} -
{\hat{h}^a}_{i-1/2} \right)
\end{equation}
where the numerical flux ${\hat{h}^a}_{i+1/2}$ is obtained with
the procedure given below. We can use the upwind fluxes without
flux splitting in the fifth order WENO approximation because the
wind direction is fixed (positive).
To obtain the sharp resolution of the contact
discontinuities, the anti-diffusive flux corrections
are used in our code.

First, we denote
$$
{h_i} = J(t^n, r_i, \nu_j), \qquad i=-2, -1, ..., {N_r}+2
$$
where $n$ and $j$ are fixed.
The numerical flux from the regular WENO procedure is obtained by
$$
\hat{h}_{i+1/2}^{-} = \omega_1 \hat{h}_{i+1/2}^{(1)}
+ \omega_2 \hat{h}_{i+1/2}^{(2)} + \omega_3 \hat{h}_{i+1/2}^{(3)}
$$
where $\hat{h}_{i+1/2}^{(m)}$ are the three third order fluxes on
three different stencils given by
\begin{eqnarray*}
\hat{h}_{i+1/2}^{(1)} & = &
\frac{1}{3} h_{i-2} - \frac{7}{6} h_{i-1}
               + \frac{11}{6} h_{i}, \\
\hat{h}_{i+1/2}^{(2)} & = &
-\frac{1}{6} h_{i-1} + \frac{5}{6} h_{i}
               + \frac{1}{3} h_{i+1}, \\
\hat{h}_{i+1/2}^{(3)} & = &
\frac{1}{3} h_{i} + \frac{5}{6} h_{i+1}
               - \frac{1}{6} h_{i+2},
\end{eqnarray*}
and the nonlinear weights $\omega_m$ are given by
$$
\omega_m = \frac {\tilde{\omega}_m}
{\sum_{l=1}^3 \tilde{\omega}_l},\qquad
 \tilde{\omega}_l = \frac {\gamma_l}{(\varepsilon + \beta_l)^2} ,
$$
with the linear weights $\gamma_l$ given by
$$
\gamma_1=\frac{1}{10}, \qquad \gamma_2=\frac{3}{5},
\qquad \gamma_3=\frac{3}{10},
$$
and the smoothness indicators $\beta_l$ given by
\begin{eqnarray*}
\beta_1 & = & \frac{13}{12} \left( h_{i-2} - 2 h_{i-1}
                             + h_{i} \right)^2 +
         \frac{1}{4} \left( h_{i-2} - 4 h_{i-1}
                             + 3 h_{i} \right)^2  \\
\beta_2 & = & \frac{13}{12} \left( h_{i-1} - 2 h_{i}
                             + h_{i+1} \right)^2 +
         \frac{1}{4} \left( h_{i-1}
                             -  h_{i+1} \right)^2  \\
\beta_3 & = & \frac{13}{12} \left( h_{i} - 2 h_{i+1}
                             + h_{i+2} \right)^2 +
         \frac{1}{4} \left( 3 h_{i} - 4 h_{i+1}
                             + h_{i+2} \right)^2 .
\end{eqnarray*}
$\varepsilon$ is a parameter to avoid the denominator
to become 0 and is taken as $\varepsilon = 10^{-5}$ times
the maximum magnitude of the initial condition $J$ in the
computation of this paper. The reconstruction of the finite difference
WENO flux on the downwind side $\hat{h}_{i+1/2}^{+}$ is obtained
in a mirror symmetric fashion with respect to $x_{i+1/2}$ as that
for $\hat{h}_{i+1/2}^{-}$.

The anti-diffusive flux corrections are based on the fluxes
obtained from the regular WENO procedure. It is given by
\begin{eqnarray}
\label{eq41}
{\hat{h}^a}_{i+1/2} & = &
               \hat{h}_{i+1/2}^{-} + \\ \nonumber
 & &  \phi_{i}
{\rm minmod}\left (\frac{{h_i} - {h_{i-1}}}{\eta} + \hat{h}_{i-1/2}^{-} -
    \hat{h}_{i+1/2}^{-},
\hat{h}_{i+1/2}^{+}-\hat{h}_{i+1/2}^{-} \right ),
\end{eqnarray}
where $\eta= \triangle t/ \triangle r$ is the CFL number
and the minmod function is defined as
\begin{equation}
   {\rm minmod}(a,b)=\left\{\begin{array} {l}
        \displaystyle    0,\hspace{1cm} if\hspace{0.5cm} ab\leq 0 \\
        \displaystyle    a,\hspace{1cm}
  if\hspace{0.5cm} ab > 0, \mid a\mid \leq \mid b \mid \\
        \displaystyle    b,\hspace{1cm}
   if\hspace{0.5cm} ab > 0, \mid b\mid < \mid a \mid . \\
         \end{array}
   \right.
\end{equation}
$\phi_{i}$ in eq.(\ref{eq41}) is the discontinuity indicator
between 0 and 1, defined as
$$
\phi_i = \frac{\beta_i}{\beta_i + \gamma_i},
$$
where
$$
\beta_i  = \left( \frac{\alpha_i}{\alpha_{i-1}}+
   \frac{\alpha_{i+1}}{\alpha_{i+2}} \right)^2, \qquad
\gamma_i = \frac{\mid {u_{\max}} - {u_{\min}}\mid^2}{\alpha_i}, \qquad
\alpha_i = (\mid h_{i-1}- h_i \mid + \zeta)^2,
$$
with $\zeta$ being a small positive number taken as $10^{-6}$
in our computation. $u_{\max}$ and $u_{\min}$
are the maximum and minimum values of $h_i$ for all grid points.
With the definition above, we will have
$0 \leq \phi_i \leq 1$. $\phi_i = O(\Delta r ^2)$ in the smooth
regions and $\phi_ i$ is close to 1 near strong
discontinuities.
The purpose of the anti-diffusive flux corrections is to improve
the resolution of contact discontinuities without sacrificing
accuracy and stability of the original WENO scheme.

\subsection{High order numerical integration in $\nu$}

The integration in equations (\ref{eq10}) and (\ref{eq1112}) is
approximated by
a fourth order quadrature formula
\begin{equation}
{\int_{\nu_0}^{\infty}}f(x)dx = \Delta x
{\sum_{j=j_0}^{\infty}}{w_j}f(j \Delta x) + O( \Delta x^4)
\end{equation}
where $\nu_0 = j_0 \Delta x$, and the weights $w_j$ are given by
$$
{w_{j_0}}=\frac{3}{8},  \quad {w_{j_0+1}}=\frac{7}{6}, \quad {w_{j_0+2}}
=\frac{23}{24},\quad {w_{j_0+j}}=1, \qquad {\rm for} \ \  j > 2.
$$
Again we refer to (Qiu et al. 2006) for details of implementation.

\subsection{Time integration}

When considering the time integration for the system of equations
(\ref{eq8}), (\ref{eq9}) and (\ref{eq11}),
we start with the third order TVD Runge-Kutta time discretization
in (Shu $\&$ Osher 1988).  For
the system of ODEs ${u}_t = L({u})$ this time discretization reads
\begin{eqnarray}
\label{rk1}
u^{(1)} & = & u^n + \Delta t L(u^n, t^n)  \\
\label{rk2}
u^{(2)} & = & \frac 3 4 u^n + \frac 1 4 (u^{(1)} + \Delta t L(u^{(1)})) \\
\label{rk3}
u^{n+1} & = & \frac 13 u^n + \frac 23 (u^{(2)}+\Delta t L(u^{(2)}))
\end{eqnarray}

The Runge-Kutta method needs to be modified considering
the modification on the anti-diffusive flux $\hat{f}^a$ by
\begin{eqnarray}
\label{rk1_a}
J^{(1)} & = & J^n + \Delta t L(J^n, t^n)  \\
\label{rk2_a}
J^{(2)} & = & J^n + \frac{1}{4} \Delta t  L'(J^{n}) +
\frac{1}{4} \Delta t L(J^{(1)}) \\
\label{rk3_a}
J^{n+1} & = & J^n + \frac{1}{6} \Delta t L''(J^{n}) +
  \frac{1}{6} \Delta t L(J^{(1)})+ \frac{2}{3}\Delta t L(J^{(2)})
\end{eqnarray}
where the spatial derivative ${\partial J}/{\partial r}$ term in operator
$L$ is defined by (\ref{shuadd60}) with the anti-diffusive flux
$\hat{h}^a$ given by eq.(\ref{eq41}), and ${\partial J}/{\partial r}$
in the operator $L'$ is defined by the modified anti-diffusive flux
$\overline{h}^a$ as
\begin{equation}
\label{eq61}
   \overline{h}_{i+1/2}^a=\left\{\begin{array} {l}
        \displaystyle    \hat{h}_{i+1/2}^{-}+
  {\rm minmod}\left (\frac{4(h_i-h_{i-1})}{\eta} \right .\\
   \displaystyle \hspace{0.3cm}
            \left . +\hat{h}_{i-1/2}^{-}-\hat{h}_{i+1/2}^{-},
                         \hat{h}_{i+1/2}^{+}-\hat{h}_{i+1/2}^{-}\right ),
                     \\
    \displaystyle \hspace{3cm} if\hspace{0.1cm} bc>0, \mid b\mid
         <\mid c\mid, \\
   \displaystyle    \hat{h}_{i+1/2}^a,\hspace{2cm} {\rm otherwise}\\
      \end{array}
   \right.
\end{equation}
and ${\partial J}/{\partial r}$ in the operator $L''$ is defined by the modified anti-diffusive flux
$\widetilde{h}^a$,
\begin{equation}
\label{eq71} \widetilde{h}_{i+1/2}^a=\left\{\begin{array} {l}
        \displaystyle    \hat{h}_{i+1/2}^{-}
      + {\rm minmod}\left (\frac{6(h_i-h_{i-1})}{\eta} \right .\\
\displaystyle  \hspace{0.3cm}
      \left . +\hat{h}_{i-1/2}^{-}-\hat{h}_{i+1/2}^{-},
                         \hat{h}_{i+1/2}^{+}-\hat{h}_{i+1/2}^{-} \right ), \\
\displaystyle \hspace{3cm} if\hspace{0.5cm} bc>0,
     \mid b\mid <\mid c\mid, \\
        \displaystyle    \hat{h}_{i+1/2}^a,\hspace{1cm} {\rm otherwise}\\
         \end{array}
   \right.
\end{equation}
Here $b =(h_i-h_{i-1})/\eta +
{\hat{h}_{i-1/2}^{-}} - \hat{h}_{i+1/2}^{-}$, $c = {\hat{h}_{i+1/2}^{+}} -
\hat{h}_{i+1/2}^{-}$.

The difficulty of a direct implementation of the scheme lies in the
stiffness of equations
(\ref{eq9}) and (\ref{eq11}). Especially for strong sources, when J(r=0)
is large, one
needs very small time step $\Delta t$ to guarantee the stability of the
numerical scheme,
therefore huge computational cost for long time integration.

By observing that
\begin{itemize}
\item Though (\ref{eq9}) and (\ref{eq11}) are stiff ODEs, (\ref{eq8})
involving the spatial
derivative is not a stiff equation.
\item The WENO procedure with the anti-diffusive flux corrections in
approximating the
spatial derivative is the major cost at each time step.
\item Implicit numerical method in time evolution has milder time step
restriction than
the explicit one.
\end{itemize}
we have settled down with the following strategies to save the
computational cost
\begin{itemize}
\item We use different time scales to solve J in (\ref{eq8}) and to
solve $f_{HI}$ and T
in (\ref{eq9}) and (\ref{eq11}). The time step
for J, say $\Delta t$, is larger, while the time step for $f_{HI}$ and
T, say
$\delta t = \frac{\Delta t}{N_t}$ with $N_t$ being the
number of small time steps in a large time step, is much smaller. By
doing this, we
take advantage of the non-stiffness of
(\ref{eq8}) and eliminate the cost of the WENO procedure with flux
corrections for each
small time step.
\item In each small time step $\delta t$, we use a semi-implicit numerical
method for
strong sources to proceed in
time for $f_{HI}$ and T, which greatly releases the severe time step
restriction for
the sake of the stability of the scheme.
The number of small time steps $\delta t$ in one large time step
$\Delta t$ is
thus greatly reduced, hence the saving of the computational cost.
\end{itemize}
More precisely, we modify the third order TVD Runge-Kutta time
discretization in the
following procedure
\begin{enumerate}
\item We have the initial condition of J, $f_{HI}$ and $T$ at t=0.
Let us denote
$$
J^0 = \left(J^0_{i, j}\right)_{N_r \times N_\nu} =
(J(x_i, \nu_j, t=0))_{N_r \times N_\nu},
$$
$$
f^0_{HI} = (f^0_{HIi})_{N_r \times 1} =
(f^0_{HI}(x_i, t=0))_{N_r \times 1}; \quad
T^0 = (T^0_i)_{N_r \times 1}
= \left(T(x_i, t=0)\right)_{N_r \times 1} .
$$
\item For a large time step $\Delta t$, we evolve $J^n$,
$f^n_{HI}$ and $T^n$ by three
inner stages
to  $J^{n+1}$, $f^{n+1}_{HI}$ and $T^{n+1}$.
\begin{itemize}
\item $J^{n+1}$:\\
$J^{(1)}$, $J^{(2)}$, $J^{n+1}$ are updated according to (\ref{rk1_a}),
(\ref{rk2_a}), (\ref{rk3_a}).
\item $f^{n+1}_{HI}$, $T^{n+1}$: \\
Let us denote
$$
\bar{h} = (h_1, h_2) = (f_{HI}, T)
$$
and
\begin{eqnarray*}
\bar{L}(J, f_{HI}, T) &=& (L_1(J, f_{HI}, T),\ L_2(J, f_{HI}, T)) \\
&=& \left( \frac{1}{c \sigma_0} \{ \ rhs \ of \ (\ref{eq9}) \}, \
\frac{1}{c \sigma_0 k_B}
\{ \ rhs \ of \ (\ref{eq11}) \}\right)
\end{eqnarray*}
Instead of (\ref{rk1}), $f^{(1)}_{HI}$ and $T^{(1)}$ are evolved by
$$(f^{(1)}_{HI}, T^{(1)}) = \bar{h}_{(N_t)}$$
where $\bar{h}_{(N_t)}$ is the evolution of the solution by the
Euler-forward method for $N_t$ times,
with the Euler-forward method at each time step defined as
\begin{equation}
\label{euler-for}
\bar{h}_{(j+1)} = \bar{h}_{(j)} + \delta t \bar{L}(J^n, \bar{h}_{(j)});
\end{equation}
Instead of (\ref{rk2}), the second inner stage $\bar{h}^{(2)}$=$
 (f^{(2)}_{HI}, T^{(2)})$
is updated by
$$
\bar{h}^{(2)} = \frac 34 \bar{h}^n + \frac 14 \bar{h}^{(2')}
$$
with $\bar{h}^{(2')} = \bar{h}_{(N_t)}$ being the evolution of the
solution by
the Euler-forward method for $N_t$ times
based on $J^{(1)}$ as
\begin{equation}
\label{euler-for2}
\bar{h}_{(j+1)} = \bar{h}_{(j)} + \delta t \bar{L}(J^{(1)},
\bar{h}_{(j)});
\end{equation}
Finally, instead of (\ref{rk3}), $f^{n+1}_{HI}$ and $T^{n+1}$ are
computed as
$$
\bar{h}^{n+1} = \frac 13 \bar{h}^n + \frac 23 \bar{h}^{(3')}
$$
with  $\bar{h}^{(3')} = \bar{h}_{(N_t)}$ obtained in a similar
manner as in
the previous step but based on $J^{(2)}$
\begin{equation}
\label{euler-for3}
\bar{h}_{(j+1)} = \bar{h}_{(j)} + \delta t \bar{L}(J^{(2)},
\bar{h}_{(j)}) .
\end{equation}
\end{itemize}
\item
When the source is strong, (\ref{eq9}) and (\ref{eq11}) suffer from severe
time step
restriction, we use a
semi-implicit scheme instead of (\ref{euler-for}) in our code to release
the severe
time step restriction and to save computational cost. The procedure
of the implementation is the same as described above except that
(\ref{euler-for}) is replaced by
\begin{eqnarray}
\label{fhi_im}
f_{HI(j+1)} &=& f_{HI(j)} + \delta t{L_1} (J^n, f_{HI(j+1)}, T_{(j)})\\
\label{T_im}
T_{(j+1)} &=& T_{(j)} +\delta t {L_2} (J^n, f_{HI(j+1)}, T_{(j+1)})
\end{eqnarray}
(\ref{fhi_im}) is computed by solving a quadratic equation with
the root located between 0 and 1, since
$L_1$ is a quadratic function of $f_{HI}$,
and (\ref{T_im}) is computed by the Newton iteration method.
(\ref{euler-for2}) and (\ref{euler-for3}) are modified in a similar way
as (\ref{euler-for}).
\end{enumerate}

\subsection{Boundary conditions}

The boundary conditions are implemented as follows
\begin{itemize}
\item
Inflow boundary condition at r=0:

$J_{i, j}= J_{0, j}, \qquad for \qquad i = 0, -1, -2$.
\item
The boundary condition at $r=r_{max}$:

$J_{N_r+i, j} = J_{N_r-1, j} ,\qquad for \qquad i=0, 1, 2$
\end{itemize}

\subsection{Convergence study of the numerical scheme}

In this subsection, we perform a grid refinement convergence study for
the numerical scheme we proposed above, to assess the accuracy of
the computational result for the typical mesh sizes that we will use
in next section.

\begin{figure}
\centerline{
\includegraphics[width=5cm]{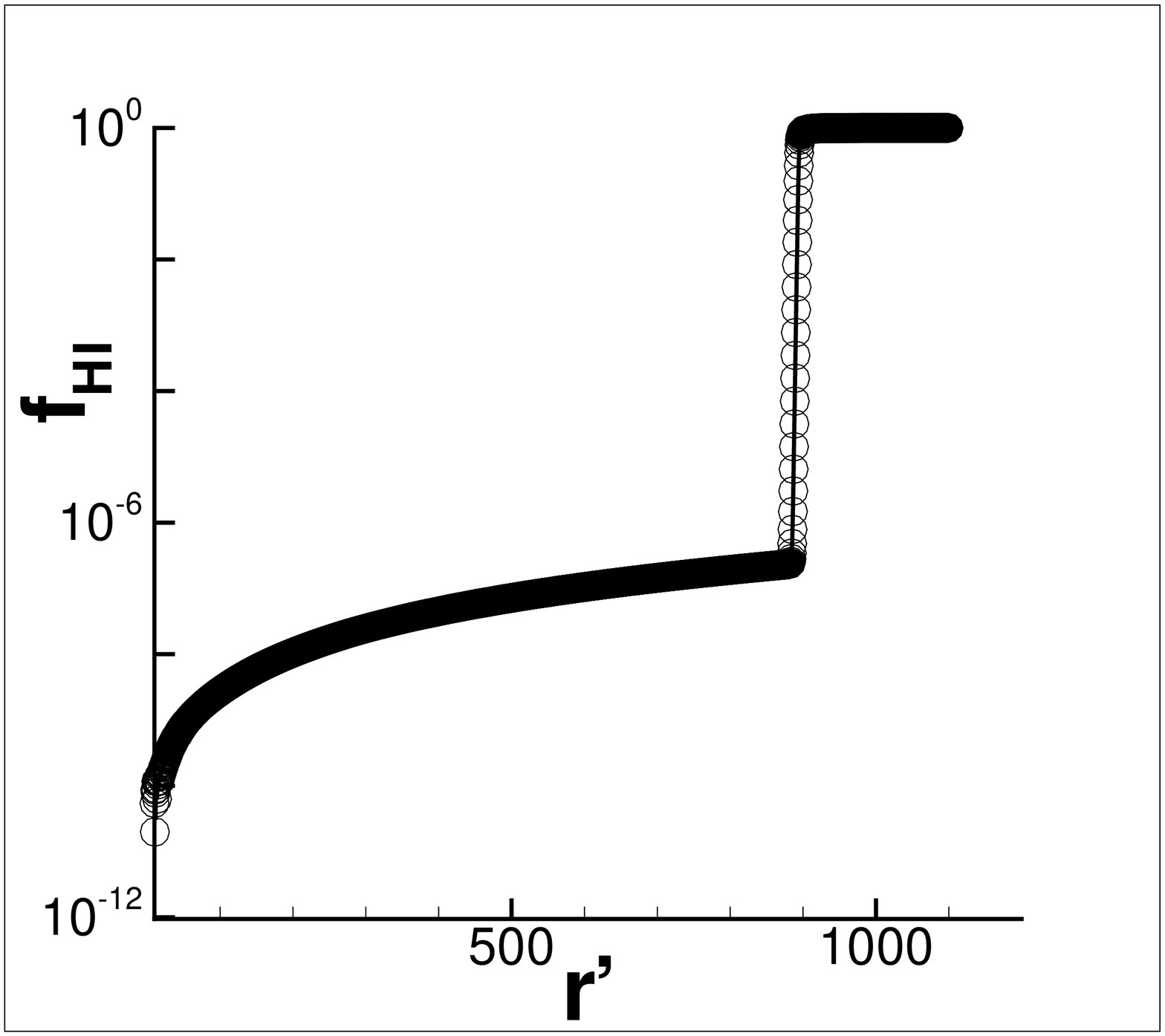}
\includegraphics[width=5cm]{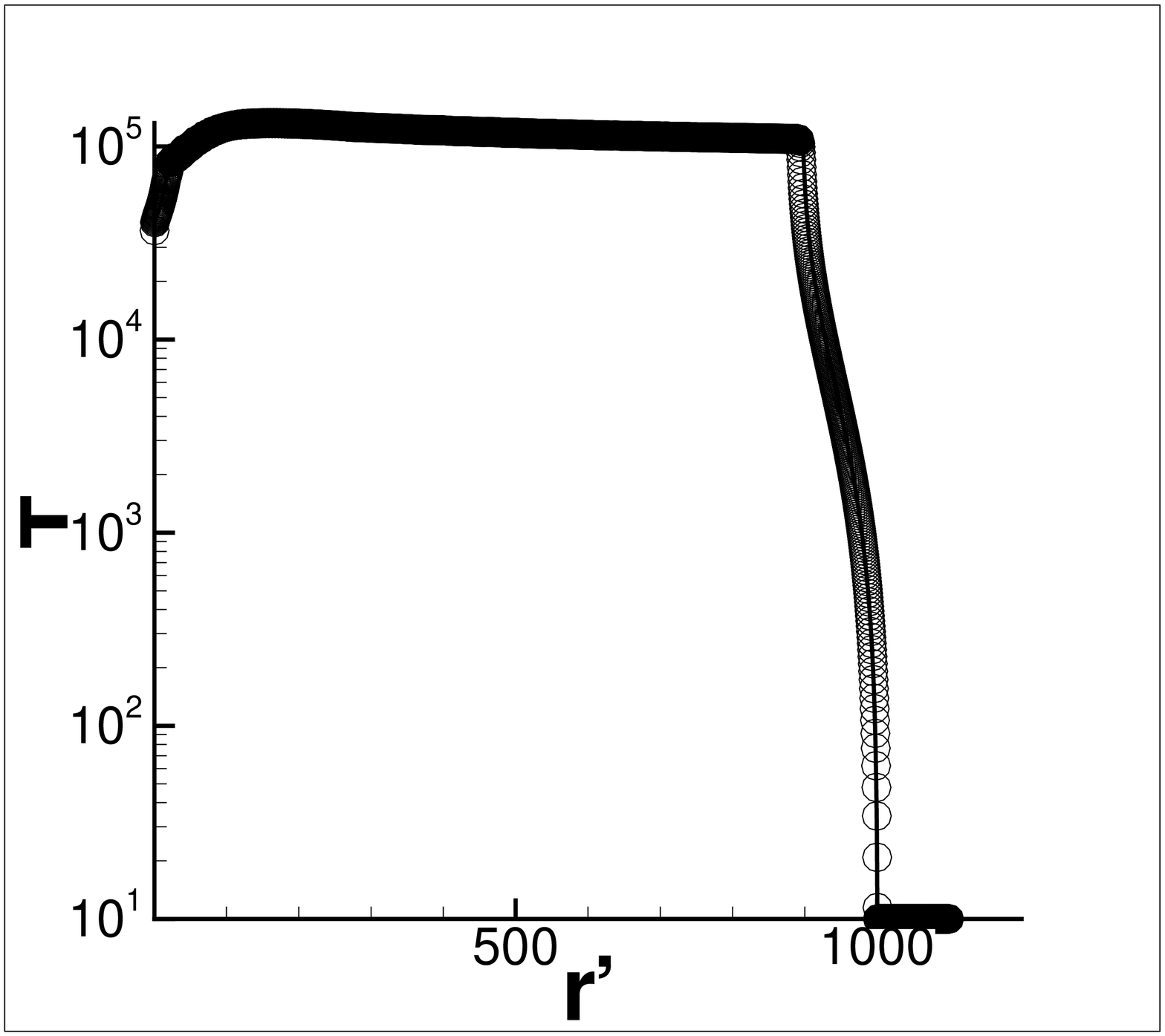}}
\caption{The profiles of $f_{\rm HI}(t,r')$ (left) and $T(t, r')$
(right) with the source of $5.8 \times 10^{45}$ erg s$^{-1}$ at time
$t=0.89$ Myrs ($t'=1000$).
The power-law frequency spectrum has the index $\alpha=2$, and the
reionization redshift is taken to be $1+z=10$.
The solid line is the numerical result with the typical mesh
$N_r = 4000$, $N_{\nu}=200$,
and the circles indicate the numerical result with a coarser mesh
$N_r = 2400$, $N_{\nu}=100$.
}
\label{fig1}
\end{figure}

We test our numerical schemes for the case with the strongest source
intensity, i.e. $\dot{E} = 5.8 \times 10^{45}$ erg $s^{-1}$. This is
the toughest situation for our numerical simulation, as the scheme
suffers severe time step restriction due to the stiffness of the
ODE.  We use both the multi-time-scale strategy and semi-implicit
time discretization described in the previous subsections, and
compare the results obtained with $4000$ points in $r$ and $200$
points in $\nu$, which is the typical mesh size used in next
section, and with a {\em coarser} mesh consisting of
$2400$ points in $r$ and $100$
points in $\nu$, in Figure \ref{fig1}.  The results with these two mesh
sizes match very well, indicating that our numerical results are
already numerically convergent at the coarser mesh.  The
numerical results reported in next section with the finer mesh
should therefore be reliable.

\begin{figure}
\centerline{
\includegraphics[width=5cm]{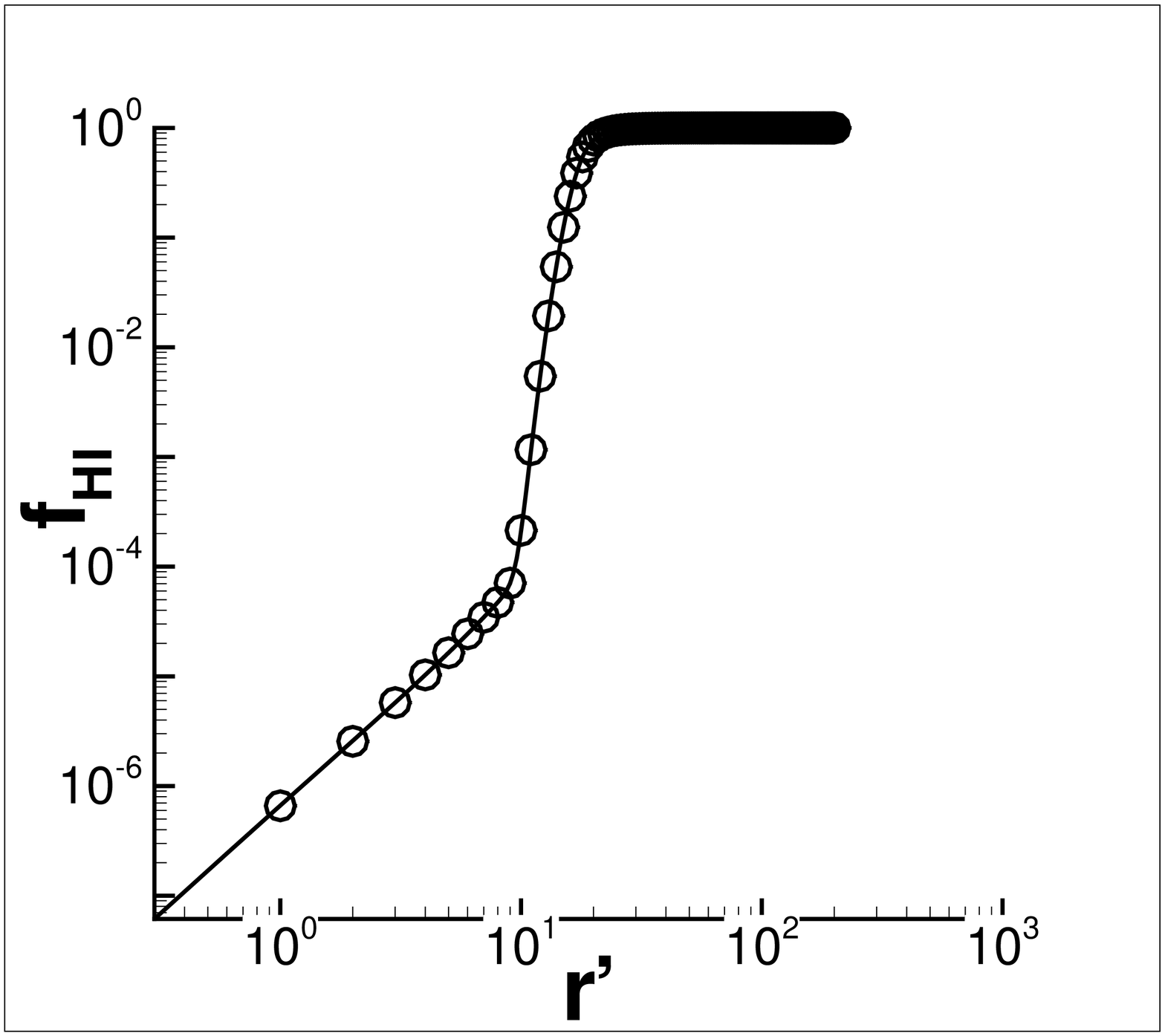}
\includegraphics[width=5cm]{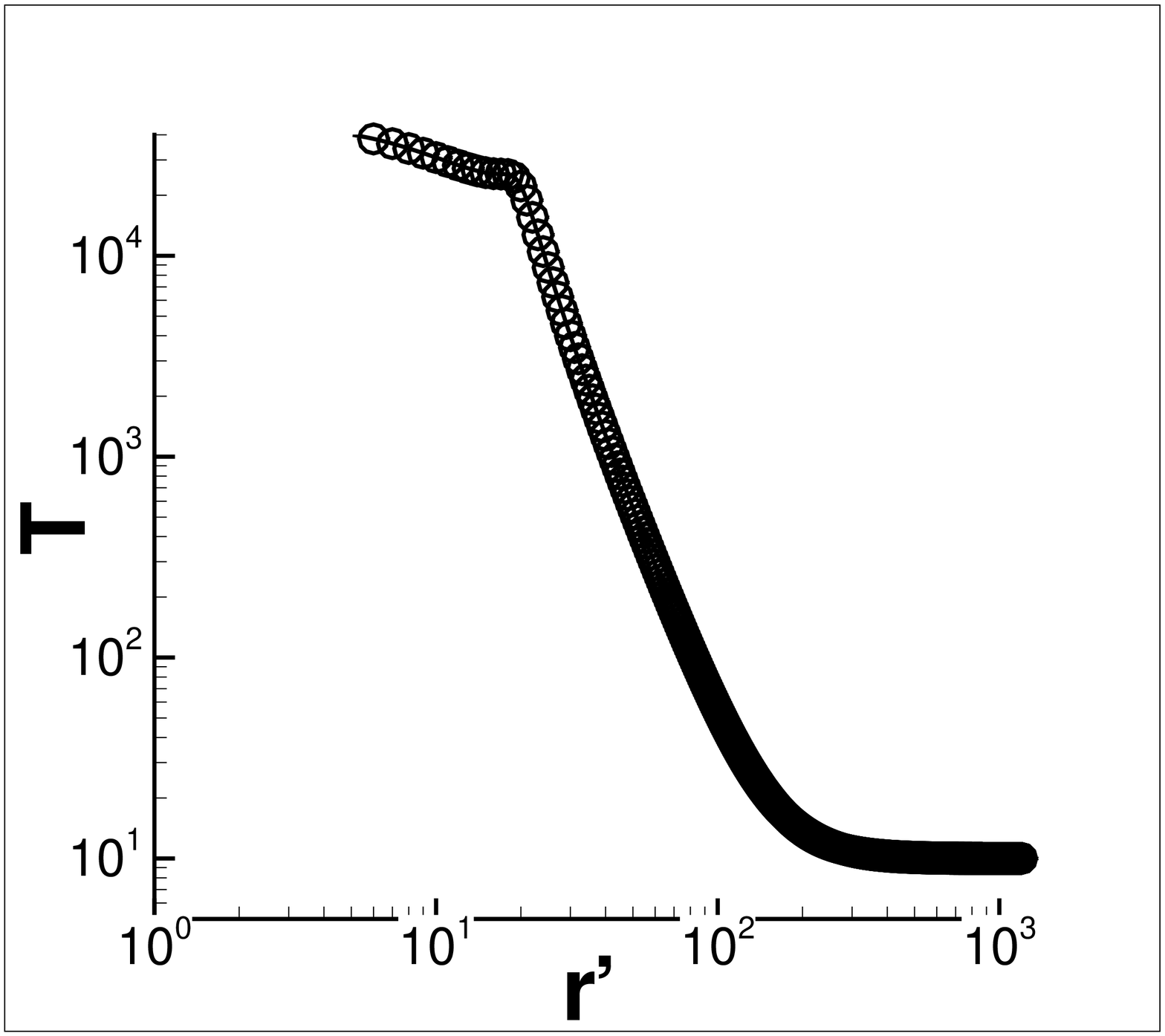}}
\caption{The profiles of $f_{\rm HI}(t,r')$ (left) and $T(t, r')$
(right) with the source of $5.8 \times 10^{39}$ erg s$^{-1}$ at time
$t=0.09$ Myrs ($t'=100$).
The power-law frequency spectrum has the index $\alpha=2$, and the
redshift is taken to be $1+z=10$.
The numerical mesh is $N_r = 2400$ and $N_{\nu}=200$.
The solid line is the numerical solution with single, extremely small time step, and
the circles indicate the numerical solution with the multi-time-scale strategy.
}
\label{fig2}
\end{figure}

We also test the accuracy of our multi-time-scale strategy by comparing
the numerical results of evolving (\ref{eq8}), (\ref{eq9}) and (\ref{eq11})
together with single, extremely small time step,
and that of updating (\ref{eq8}), (\ref{eq9}) and (\ref{eq11})
by the multi-time-scale strategy with ($\delta t$, $\Delta t$).
These two approaches produce numerical solutions in good agreement, see Figure \ref{fig2}.

Additional grid refinement study, not reported here to save space,
has been performed to assure the numerical
convergence of the results reported in next section.

\section{Results}

The results in this section are obtained with the numerical
algorithm described in the previous section.  The code is
stable for all the cases reported in this section.  As indicated in
the previous section 3.7, and we
have performed grid refinement study for some representative
examples to ensure that the results reported are numerically
converged solutions.

In our calculation, we use the explicit time discretization to
solve (\ref{eq9}) and (\ref{eq11}) when  $\dot{E} = 5.8 \times 10^{39}$ erg $s^{-1}$,
and semi-implicit scheme when  $\dot{E} = 5.8 \times 10^{41}$ erg $s^{-1}$,
$5.8 \times 10^{43}$ erg $s^{-1}$ and $5.8 \times 10^{45}$ erg $s^{-1}$
in each of the small time step $\delta t$ in the multi-time-scale strategy.

\subsection{Profiles of $f_{\rm HI}(t,r)$ and $T(t,r)$}

\begin{figure}
\centerline{
\includegraphics[width=5cm]{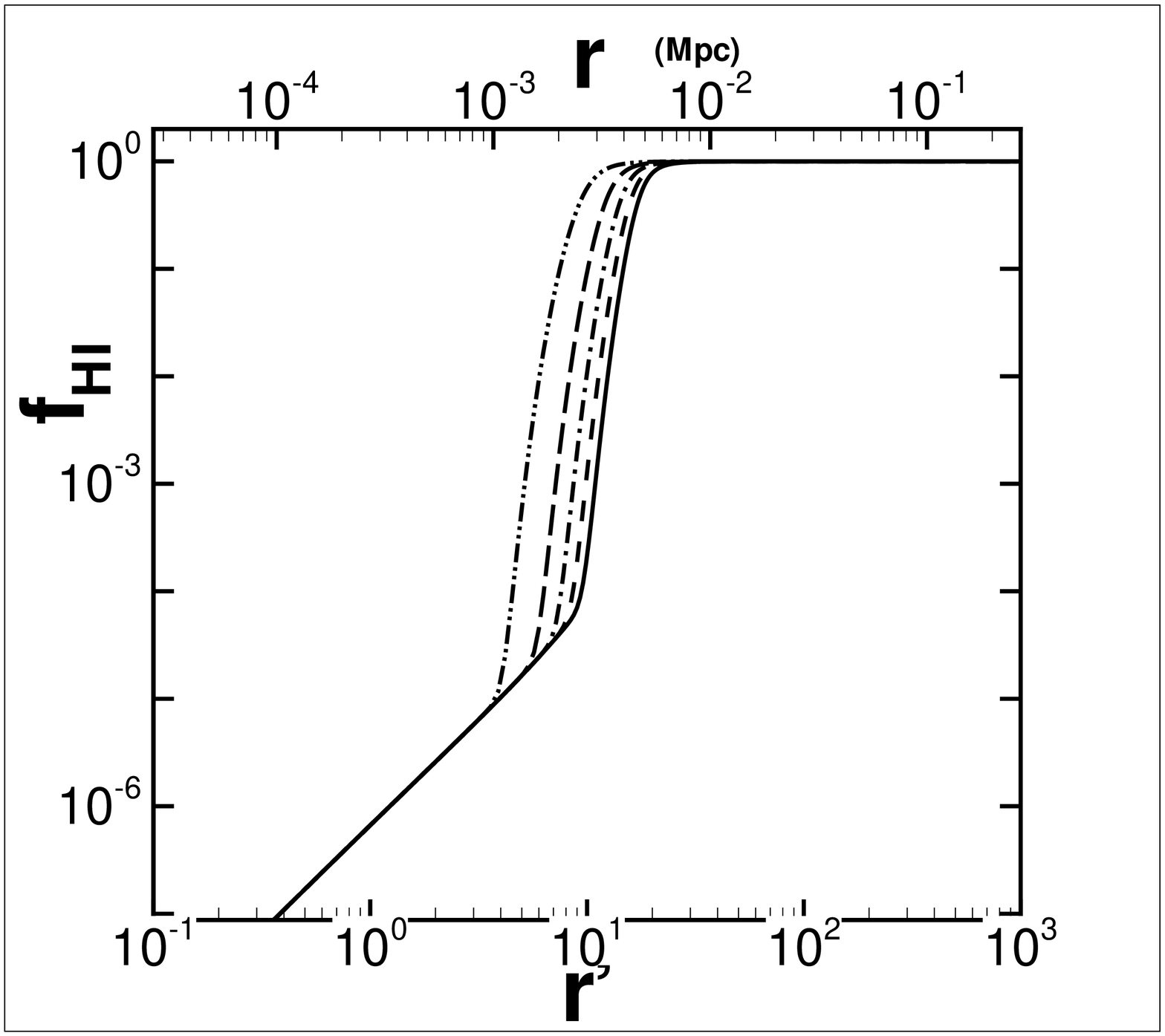}
\includegraphics[width=5cm]{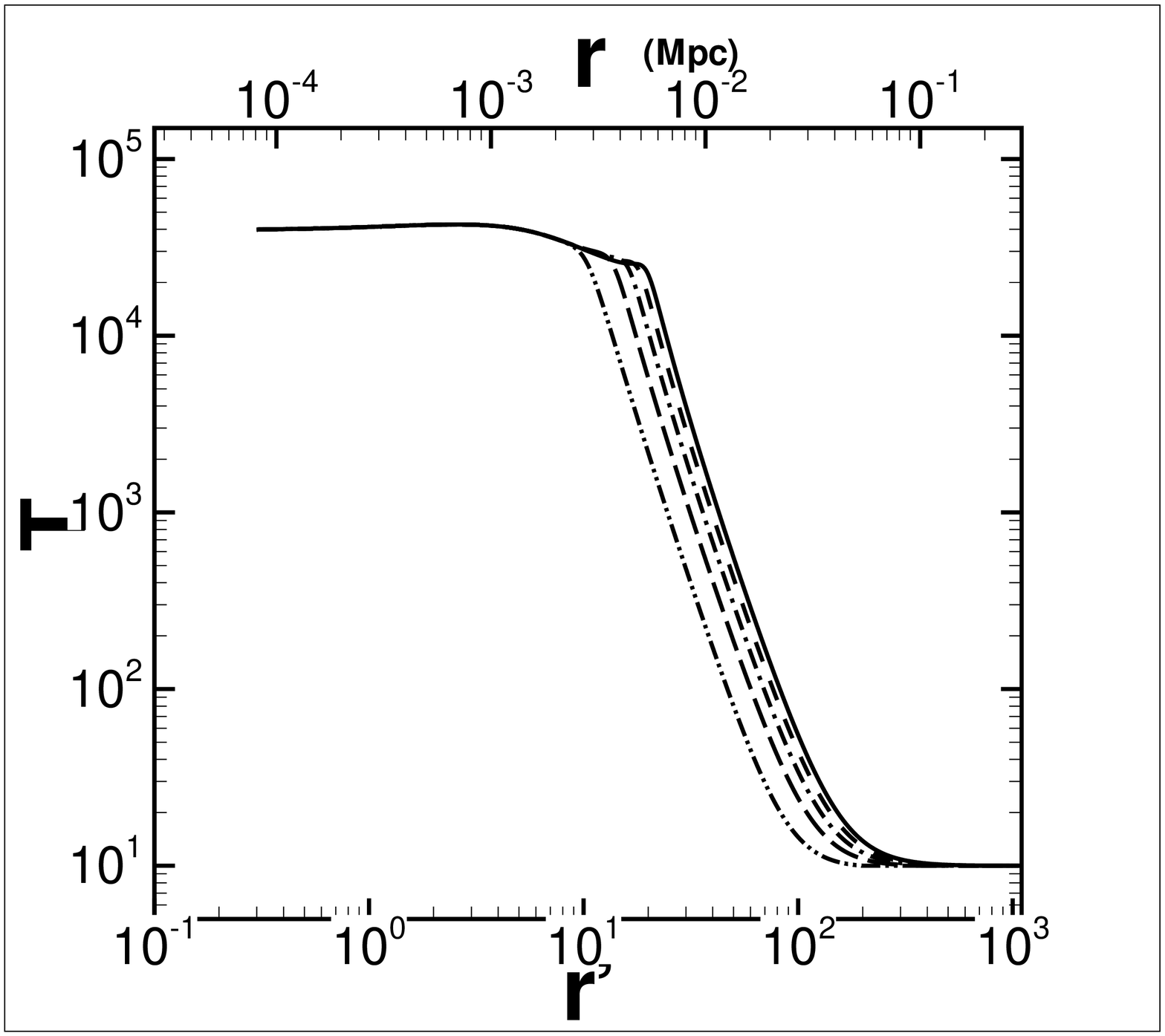}}

\vspace{0.15cm}

\centerline{
\includegraphics[width=5cm]{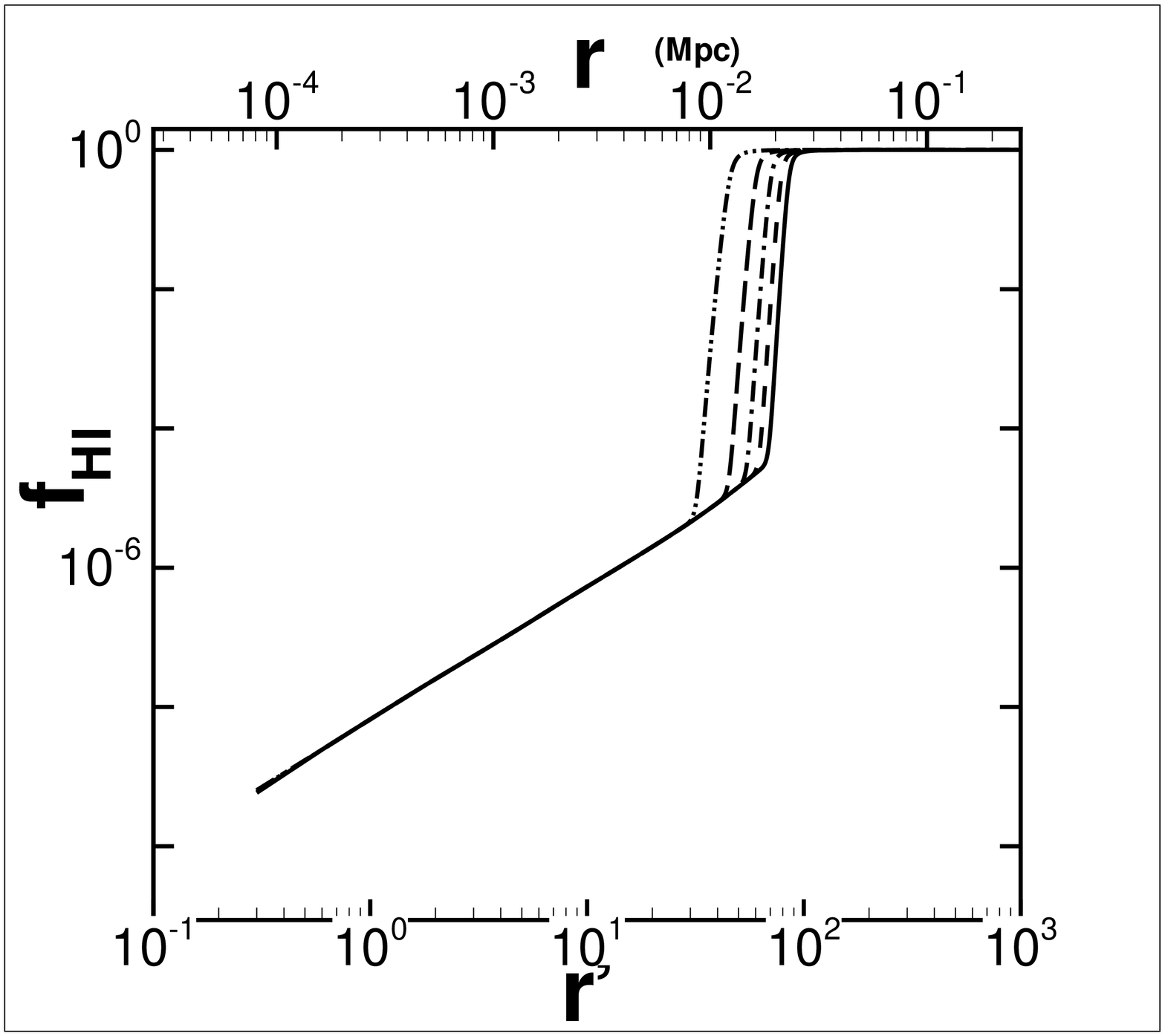}
\includegraphics[width=5cm]{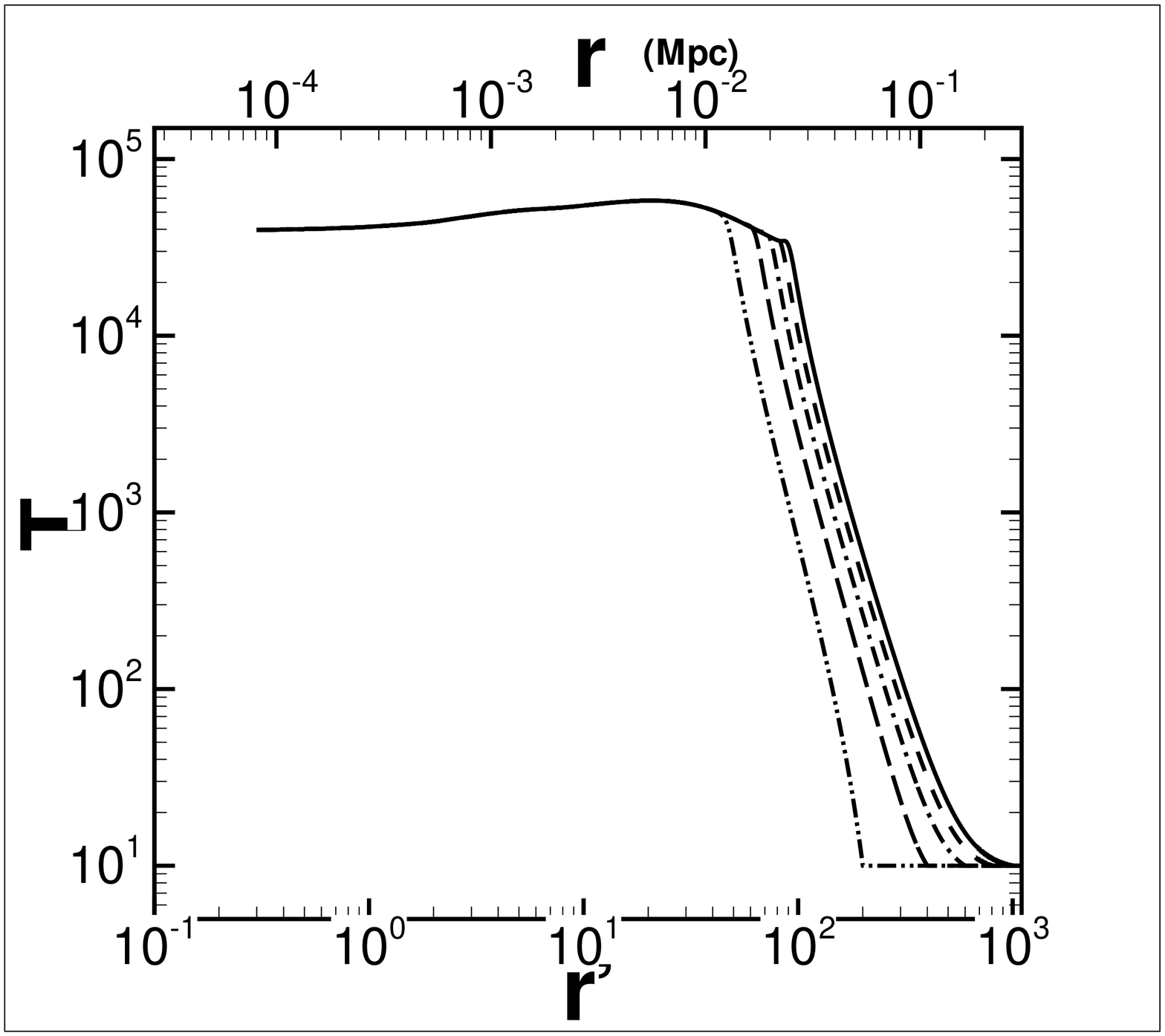}}

\vspace{0.15cm}

\centerline{
\includegraphics[width=5cm]{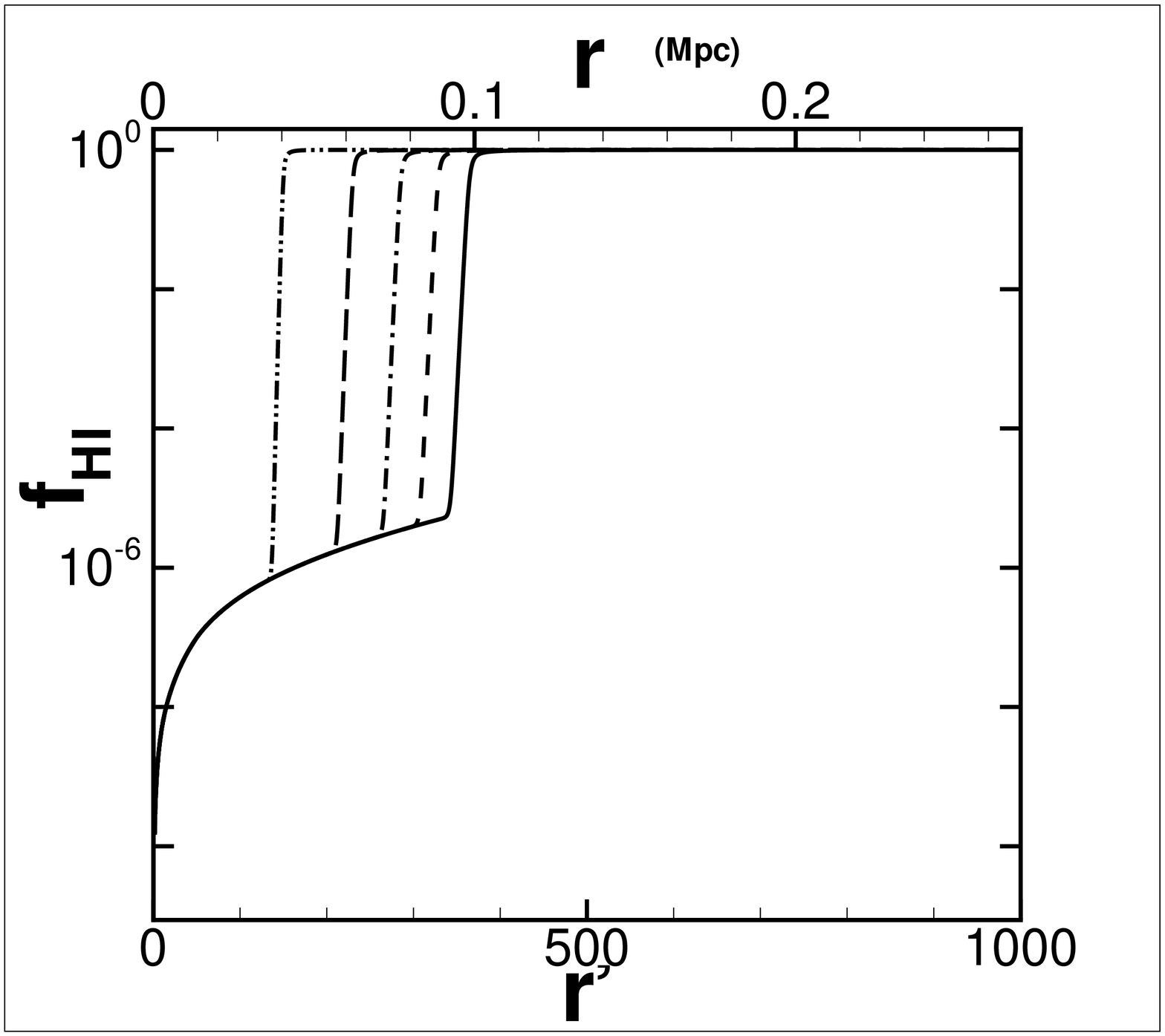}
\includegraphics[width=5cm]{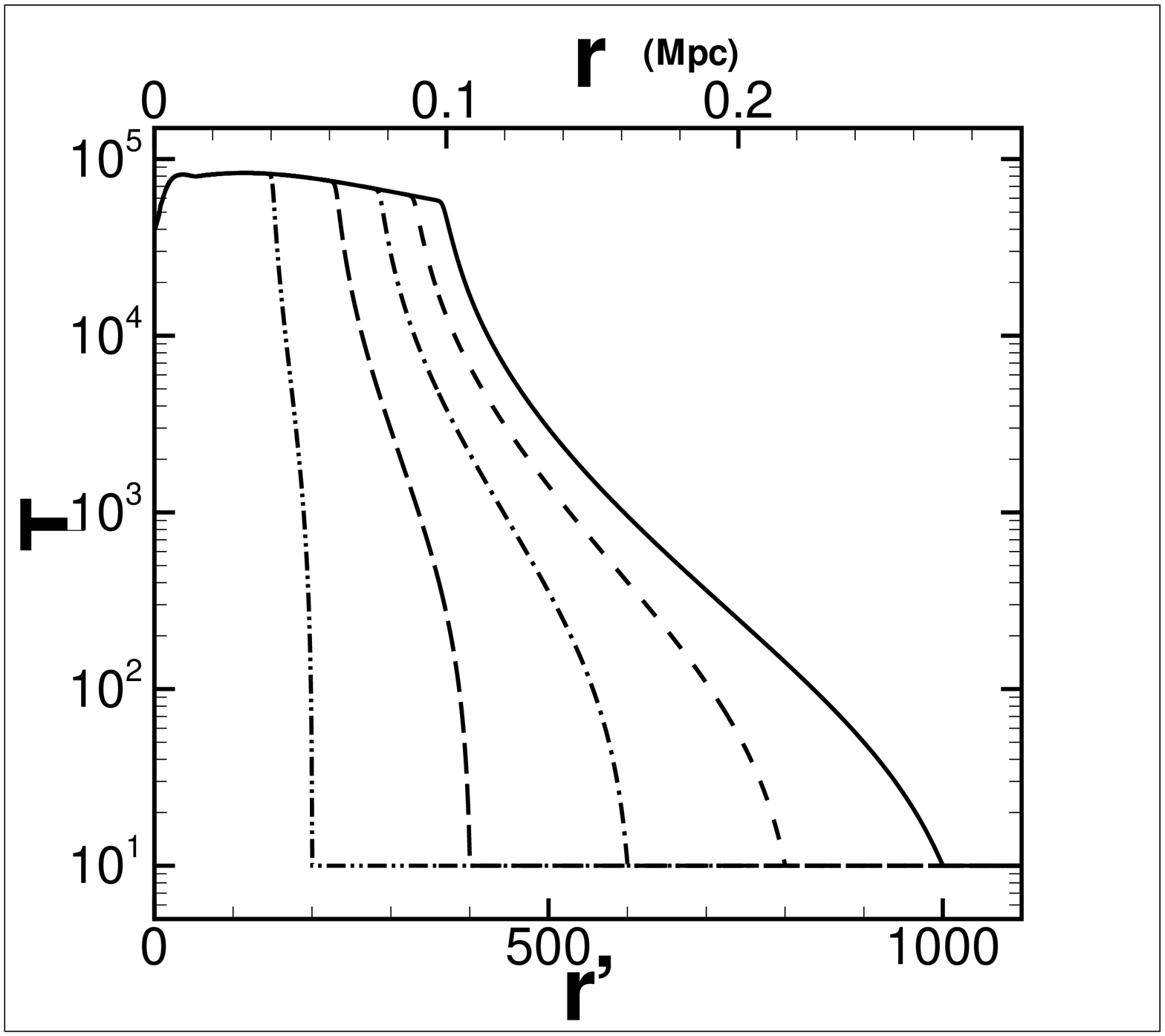}}

\vspace{0.15cm}

\centerline{
\includegraphics[width=5cm]{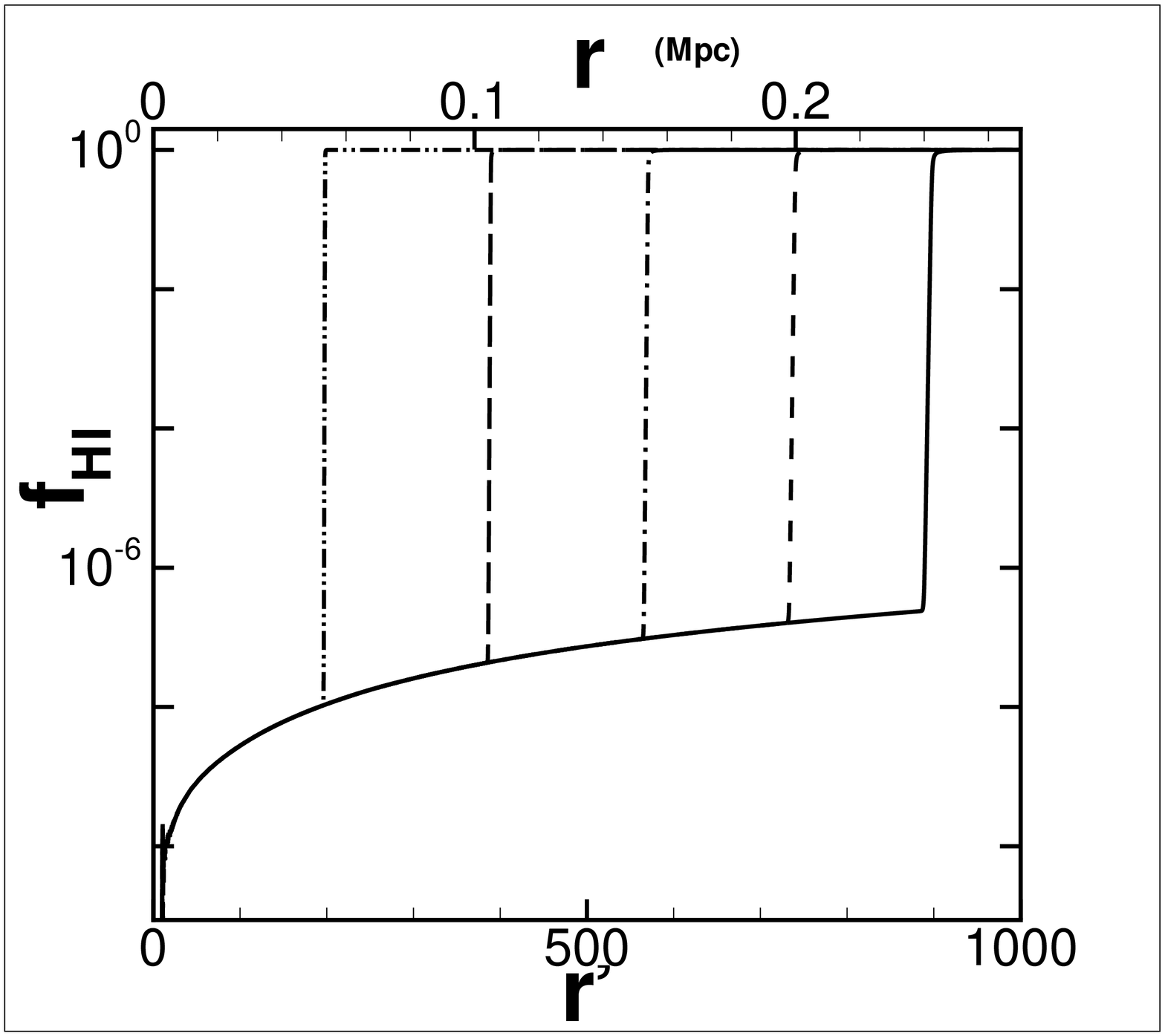}
\includegraphics[width=5cm]{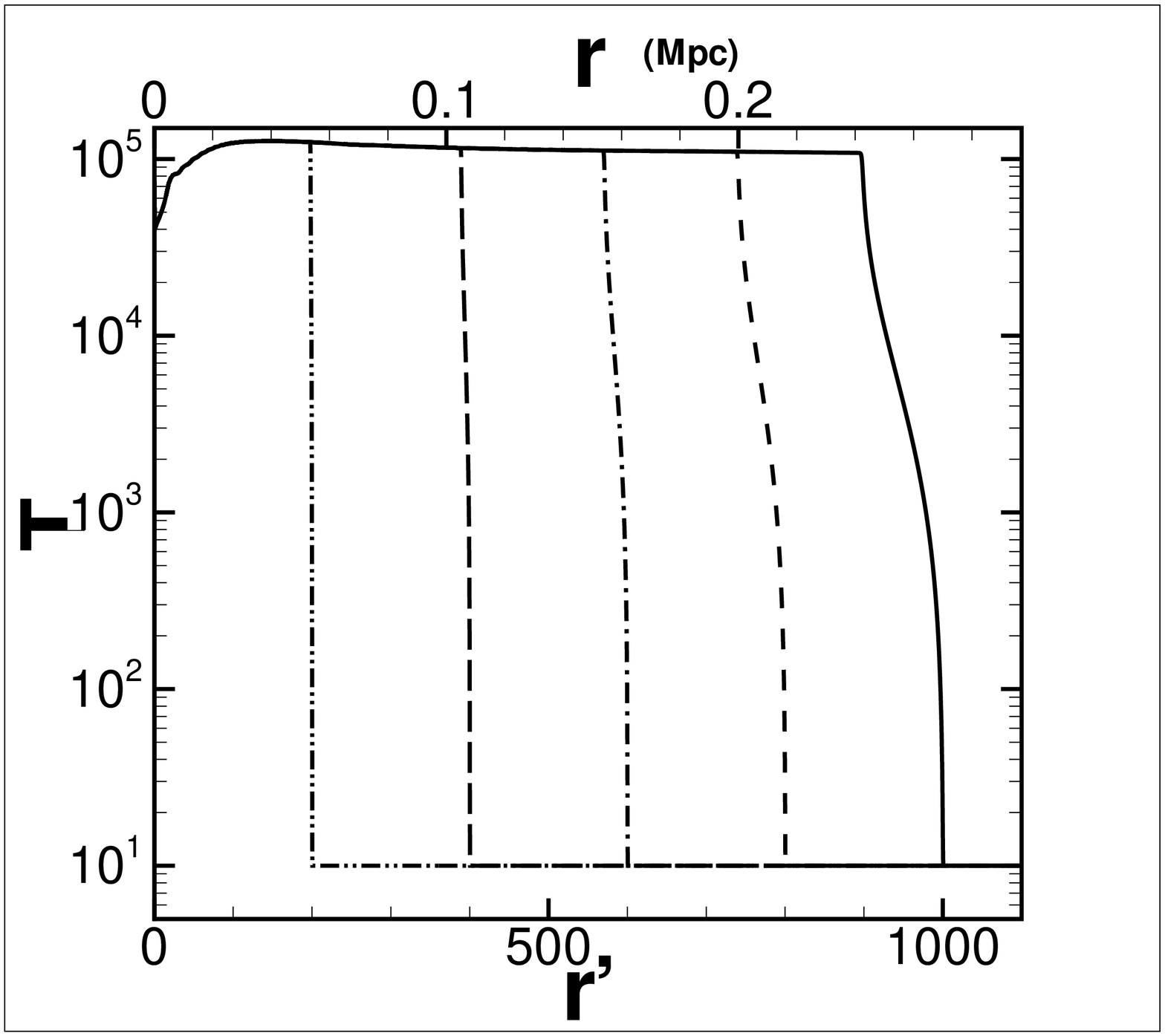}}

\caption{The profiles of $f_{\rm HI}(t,r)$ (left) and $T(t, r)$
(right), in which $r$ and $r'$ are the physical and dimenssionless
distance, respectively. From top to bottom: for sources
of  $\dot{E}=5.8 \times 10^{39}$, $5.8 \times 10^{41}$,
$5.8 \times 10^{43}$ and $5.8 \times 10^{45}$
erg s$^{-1}$ at time $t=0.18$ (dash dot dot), 0.36 (long dash),
0.53 (dash dot), 0.71 (dash), and 0.89 (solid line) Myrs. The
power-law frequency spectrum has the index $\alpha=2$, and the
redshift is taken to be $1+z=10$.  The mesh has
$N_r$=4000, $N_\nu$=200 grid points.
}
\label{fig3}
\end{figure}

We calculate the profiles of $T(t, r)$ and $f_{\rm HI}(t, r)$
around a point source. The result is presented in Figure \ref{fig3},
which shows the time-dependence
of the profiles $f_{\rm HI}(t, r)$ and $T(t, r)$ for sources with
intensity, respectively from top to bottom panels, $\dot{E}=5.8\times
10^{39}$,
$5.8\times 10^{41}$, $5.8\times 10^{43}$ and $5.8\times 10^{45}$
erg s$^{-1}$ with $\alpha=2$  and $1+z=10$.
The emission rate of the number of photons is roughly
$\dot{N}\simeq 10^{50}$, $10^{52}$, $10^{54}$ and $10^{56}$ s$^{-1}$.

Figure \ref{fig3} shows that the ionized sphere generally has a sharp
I-front. The time-dependent profile $f_{\rm HI}(t, r)$ may still be
approximately described as a Str\"omgren sphere with time-dependent
size $R(t)$ as
\begin{equation}
\label{eq22a}
f_{\rm HI}(t, r)= \left \{ \begin{array}{ll}
                       1, &  r>R(t), \\
                       \simeq 0, & r<R(t)
                      \end{array}  \right .
\end{equation}
The code can well reveal the propagation of the I-front $r=R(t)$.
Unlike the solution eq.(\ref{eq1}), $f_{\rm HI}(t,r)$ is not zero in
$r<R(t)$. Although the neutral hydrogen $f_{\rm HI}(t,r)$ remained
in $r<R(t)$ is very small, it cannot be ignored, because the heating
of gas is given by the hard photons absorption of neutral hydrogen
HI [eq.(A8)]. The high temperature within the ionized sphere is
actually maintained by the ionization of the small fraction therein
by all the ionizing photons [eq.(\ref{eq22a})]. Figure \ref{fig3}
indicates that $f_{\rm HI}(t,r)$ within $r<R(t)$ is sensitive to the
intensity of the source. This is important for investigating leakage
of the Ly-$\alpha$ photon.

Figure \ref{fig3} shows that the temperature profiles $T(t,r)$ within the
I-front $r<R(t)$ are simple.  The temperature keeps around a
constant $10^4-10^5$ K. An interesting feature is that for strong
sources, the temperature at the center ($r \approx 0$) is not the
highest and is even lower than the constant temperature. This is
because the ionizing (soft) photon flux at the center is very strong,
the number of $f_{\rm HI}(r\sim 0)$ is extremely low
(see Figure \ref{fig3}),
and then, the heating rate is lower.

A common feature of $T(t,r)$ is that the size of heated region is
generally larger than that of ionized region at all time. That is,
hydrogen gas is already significantly heated before being ionized.
The gas is firstly heated up, and then ionized. The region between
the T-front and the I-front is a pre-heating layer, in which the
temperature can be as high as $T \simeq 10^3$ K, or even larger.
For a source of $\dot{E}=10^{43}$ erg s$^{-1}$ at $(1+z)=10$, the
physical size of the ionized region at time 0.36 Mys is about
0.06 Mpc, while the physical size of the region with temperature
larger than $10^3$ K is about 0.08 Mpc. When $t=0.89$ Mys,
the two physical sizes are, respectively, about 0.1 Mpc, and 0.16
Mpc, or comoving size  1 Mpc, and 1.6 Mpc. Therefore, for strong
sources, the time scales and comoving length scales of the pre-heating
layer are in the range of cosmological interest.

An important feature of the ionization profile is that $f_{\rm
HI}(t, r)$ can be approximately written as $f_{\rm HI}(t,r)= f_{\rm
HI}(r)\theta[r-R(t)]$, where $f_{\rm HI}(r)$ is time-independent,
and $R(t)$ is only a function of $t$. That is, all the time-dependence
of $f_{\rm HI}(t,r)$ can be described by the I-front $R(t)$.
In the region $r<R(t)$, the ionization profile $f_{\rm
HI}(t,r)=f_{\rm HI}(r)$ is time-independent. It can be given by a
static or stationary solution of the radiative transfer equation.
This property has been applied in some codes for radiative transfer,
in which the ionization field is given by the static solution of the
radiative transfer equation for a given matter distribution.

For the temperature profile, one might also approximately define a
T-front
function by a step function as $\theta[t-R(t,T)]$. However, unlike the
I-front, the temperature profile can not be rewritten as
$T(t,r)=T(r)\theta[t-R(t,T)]$, where $T(r)$ is time-independent.
That is, the $r$-dependence of the function $T(t, r)$ can not be
separated with $t$. Especially, the function $T(t,r)$ in the range
between the I- and T-fronts actually always depends strongly on $t$.

\subsection{Speed of the I- and T-fronts}

\begin{figure}
\centering
\includegraphics[width=6cm]{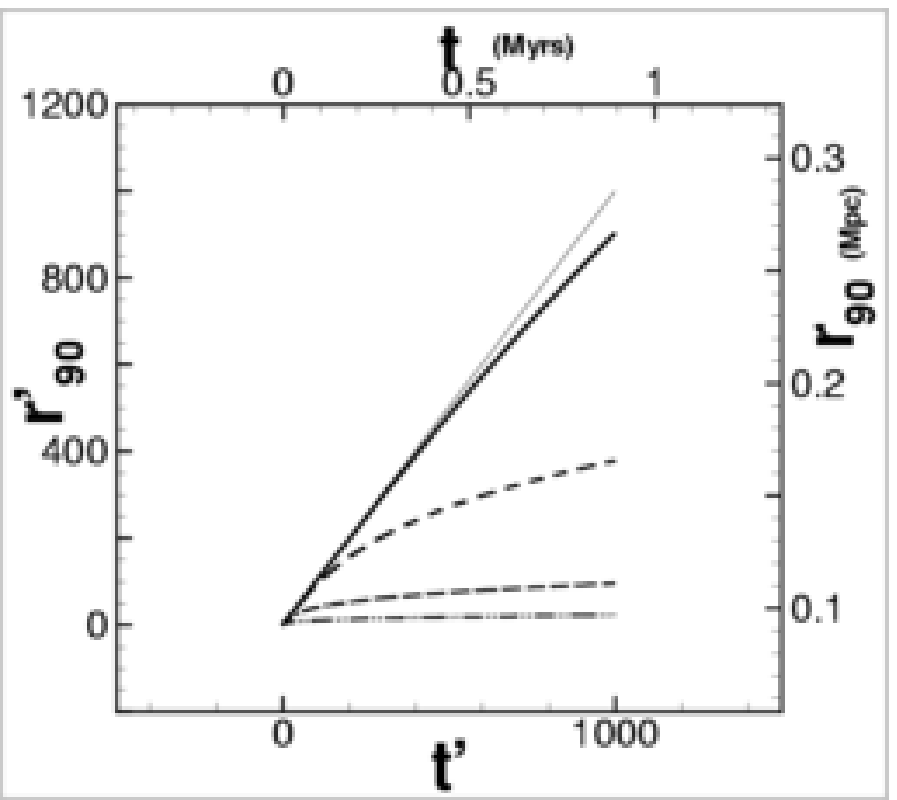}
\includegraphics[width=6cm]{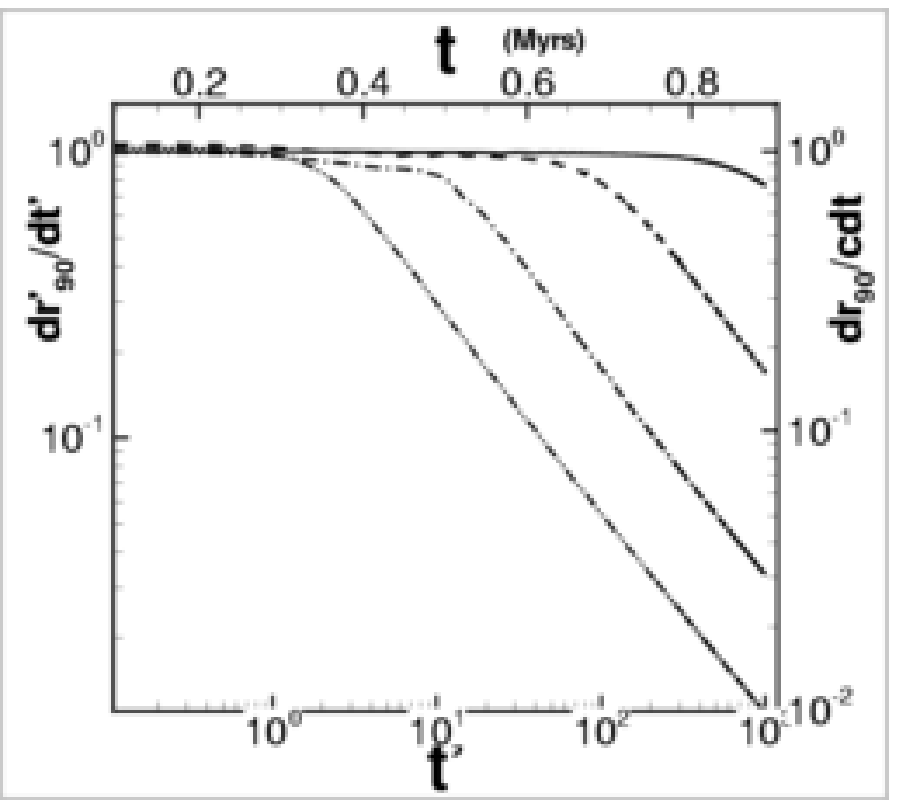}
\caption{left panel: the evolution of $r_{90}(t)$, which is the
solution of
$f_{\rm HI}(t, r_{90})=0.90$.
The light solid line is $r_{90}=ct$.
Right panel: $dr_{90}/dt$ vs. $t$.
Other parameters and the mesh in the numerical simulation
are the same as those in Figure \ref{fig3}.
The sources are taken to be
$\dot{E}=5.8\times 10^{39}$ (dash dot dot), $5.8 \times 10^{41}$
(dash dot), $5.8 \times 10^{43}$ (dash), and $5.8 \times 10^{45}$
(solid line) erg s$^{-1}$. Time $t'$ is dimensionless.
}
\label{fig4}
\end{figure}

Figure \ref{fig3} shows that, for a very strong source
$\dot{E}= 10^{45}$ erg
s$^{-1}$, we have $R(t)\simeq ct$, i.e. the ionizing front moves
with a speed close to the speed of light. For a weak source
$\dot{E}=10^{38}$ erg s$^{-1}$, the profile $f_{\rm HI}(t, r)$ does
not show a significant expansion of the I-front if time $t$ is larger
than a few of free flight times of ionizing photons. This result
is also evident in Figure \ref{fig4} on $r_{90}(t)$ versus $t$.
$r_{90}(t)$ is defined
by $f_{\rm HI}(t, r_{90})=0.90$. It denotes the size, within which,
i.e. $r<r_{90}$, 90\% hydrogen are ionized. Therefore, it can be
used for the I-front. For a strong source, such as $\dot{E}=10^{45}$
ergs s$^{-1}$, we have approximately $r_{90}=ct$, which implies that
the ionizing region grows with an ionizing front propagating with
almost the speed of light. For weak sources, $r_{90}(t)$ are also
following $t$ at very small $t$, but become $r_{90}(t)<ct$ when $t$
is large. The weaker the sources, the earlier the stage of
$r_{90}(t)<ct$ takes place. This point is more clearly shown in the
right panel of Figure \ref{fig4}. We see that the speed $dr_{90}(t)/dt$ is
close to $c$ when $t$ is small, and then the speed decreases with $t$ by
a power law $dr_{90}(t)/dt \propto t^{-\beta}$ with $\beta >2/3$.
One can define a time $t_i$, larger than which, the speed
$dr_{90}(t)/dt$ starts to decrease with $t$ by the power law. At
time $t>t_i$, the
ionizing sphere is still expanding, but very slowly. It will finally
approach a solution, of which the ionization equilibrium is
approximately established, and the ionized sphere becomes
time-independent. The time $t_i$ depends on the source intensity.
The stronger the source, the larger the time $t_i$.

Obviously, the speed of the propagation of T-front cannot be larger
than the speed of light, and therefore, the T-front will approximately
coincide
with the I-front when $t<t_i$. Only in this period, the ionized sphere
is the same as the heated sphere. When $t>t_i$, the T-front starts
to exceed the I-front, and the pre-heating layer is formed. Therefore,
the formation of pre-heated layer happens later for a stronger source.

\begin{figure}
\centering
\includegraphics[width=6cm]{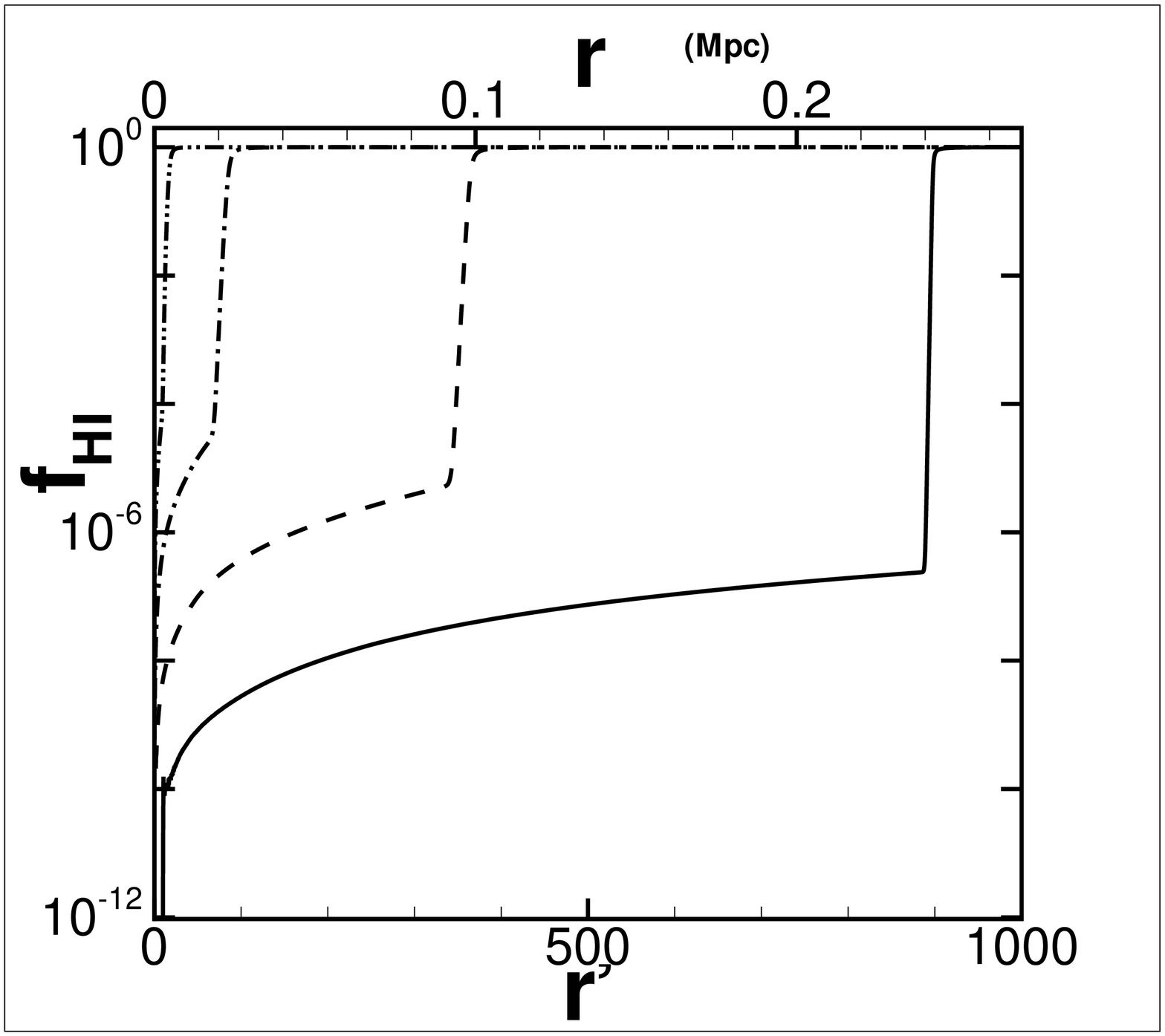}
\includegraphics[width=6cm]{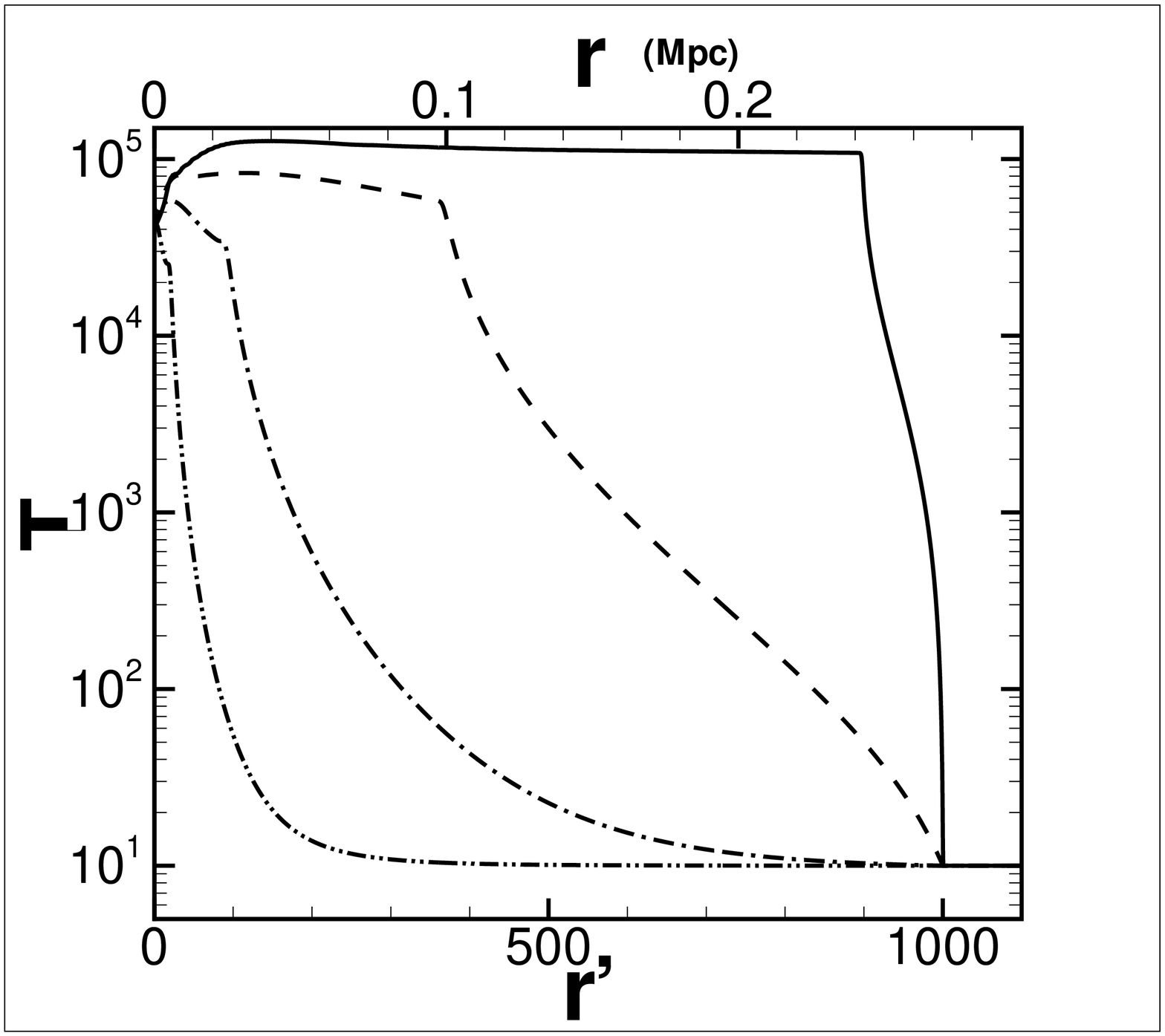}
\caption{Profiles of  $f_{\rm HI}(r)$ (left panel) and  $T(r)$
(right panel) at time $t=0.89$ Myrs for sources with intensity
$\dot{E}=5.8\times 10^{39}$ (dash dot dot), $5.8 \times 10^{41}$
(dash dot), $5.8 \times 10^{43}$ (dash), and $5.8 \times 10^{45}$
(solid line) erg s$^{-1}$. $r$ and $r'$ are the physical and
dimensionless distance, respectively. The power-law frequency
spectrum has index $\alpha=2$. The redshift is taken to be $1+z=10$.
The mesh in the numerical simulation is the same as that in Figure
\ref{fig3}. } 
\label{fig5}
\end{figure}

Figure \ref{fig5} gives a comparison of $f_{\rm HI}(t,r)$ and $T(t,r)$
at time
$t=0.89$ Myrs for sources with different intensity. We can see that
the pre-heating layer has been well established at
time $t=0.89$ Myrs for all sources with $\dot{E}\leq 10^{43}$
erg s$^{-1}$, while the T-front of the source with $\dot{E} =
10^{45}$ erg s$^{-1}$ is still about the same as the I-front.
One can expect that for a source of $\dot{E}\geq
10^{45}$ erg s$^{-1}$, a preheated layer will be formed at time
$t>0.89$ Myrs and on comoving distance $r>2.7$ Mpc from the source.

\subsection{Spectral hardening}

The formation of the pre-heated layer with high $T$ and high
$f_{\rm HI}$ is due to the lack of soft photons beyond the I-front,
while hard photons are still abundant in that region. In other words,
the energy spectrum of the photons is significantly hardened around
the I-front. We now directly demonstrate the evolution of the photon
frequency spectrum, as our code can effectively reveal the
evolution of radiations in the $\nu$-space.

\begin{figure}
\centering
\includegraphics[width=6cm]{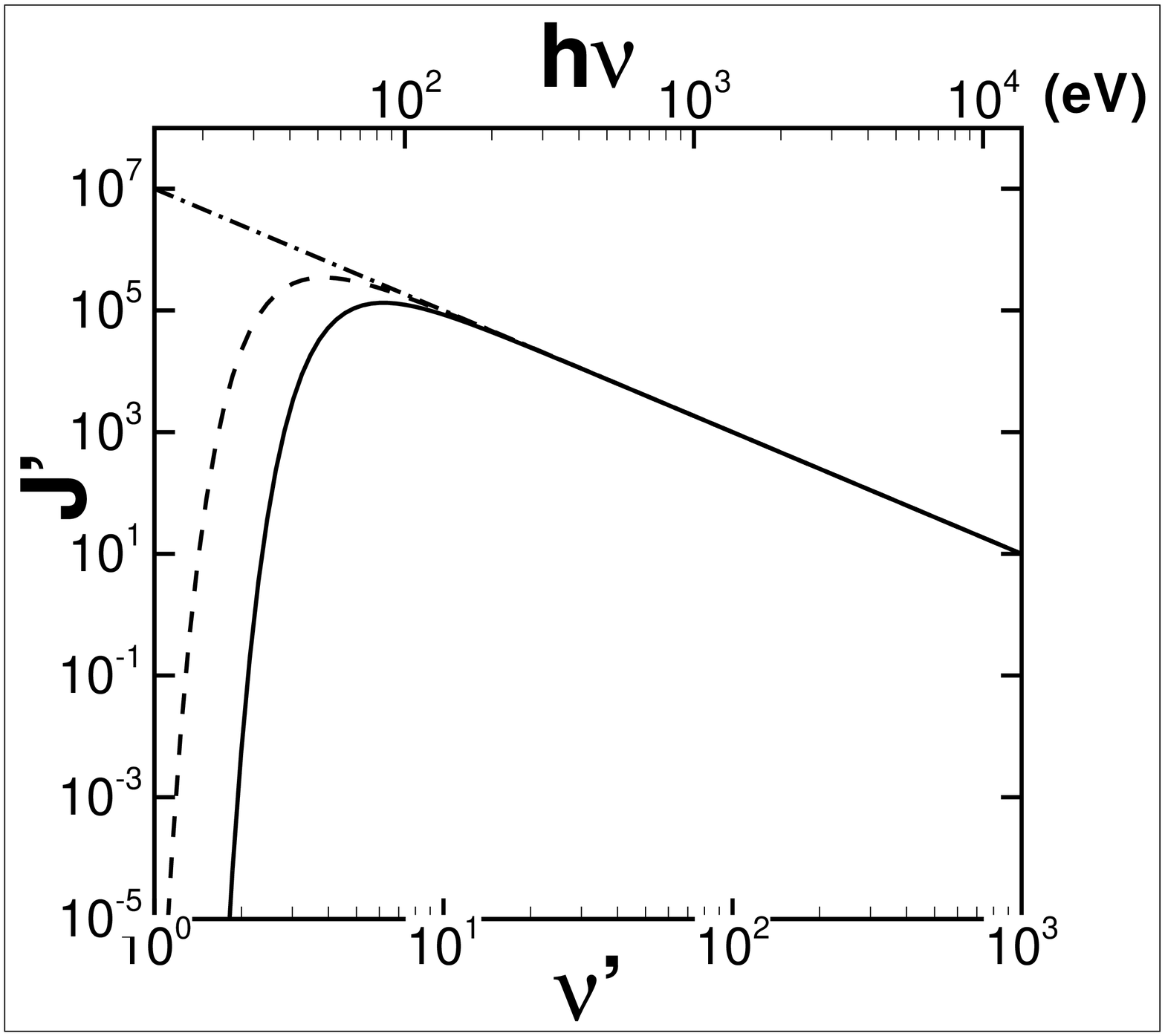}
\includegraphics[width=6cm]{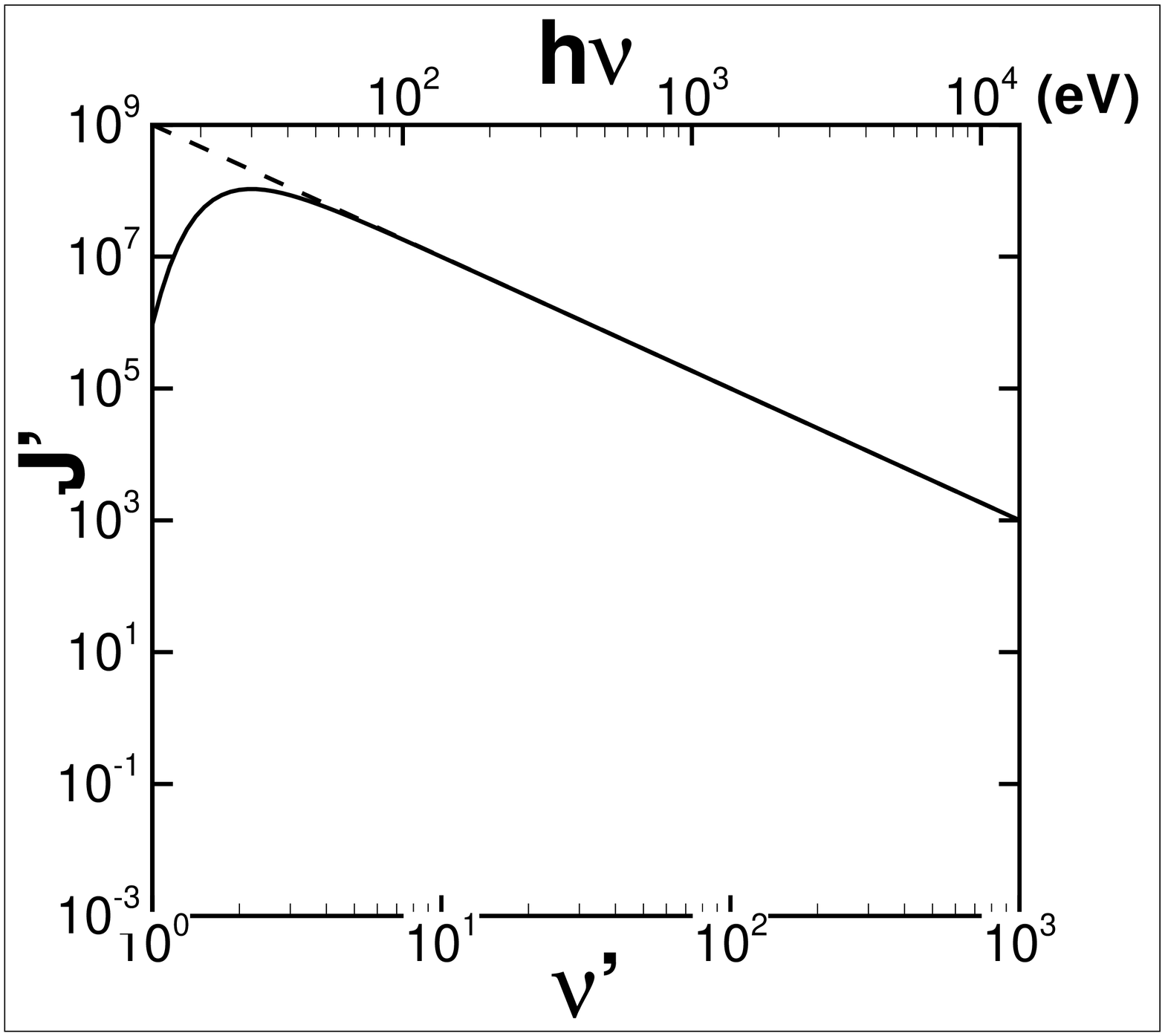}
\caption{$J'(t,r,\nu)$ vs. $\log \nu/\nu_0$ at time $t=0.89$ Myrs.
1.) Left panel: for the source $\dot{E}=5.8\times 10^{43}$ erg s$^{-1}$,
at physical distance $r=0.08$ (dash dot), 0.11
(dash) and 0.14 (solid line) Mpc; and 2.) Right panel: for the source
$\dot{E}=5.8\times 10^{45}$ erg s$^{-1}$ at physical distance
$r=0.19$ (dash) and
0.24 (solid line) Mpc. Redshift is $1+z=10$.
The mesh in the numerical simulation is the same as that
in Figure \ref{fig3}.
}
\label{fig6}
\end{figure}

The left panel of Figure \ref{fig6} gives the frequency spectra of
1.) source
$\dot{E}=5.8 \times 10^{43}$ erg s$^{-1}$ at time $t=0.89$ Myrs and
physical distance $r=0.08$, 0.11, and 0.14 Mpc, and 2.) source
$\dot{E}=5.8\times 10^{45}$ erg s$^{-1}$ at time $t=0.89$ Myrs and
physical distance $r=0.19$ and 0.24 Mpc. We can see a significant
$r$-dependence of the frequency spectrum. At a small $r$ the spectra
are still almost the same as the original power-law spectrum
$\nu^{-\alpha}$ with $\alpha=2$, while at large $r$, they significantly
deviate from the original power-law. All photons of $\nu<10\nu_0$,
are exhausted within $r < 0.24$ Mpc.  At high
frequency $\nu>50\nu_0$, the spectra are still of power-law with
$\alpha=2$, while at $\nu < 50\nu_0$ they are substantially dropped. It
shows a peak at $1 < \nu/\nu_0 <10$, and looks like a spectrum with
self-absorption. However, unlike a self-absorption spectrum, the
position of the peak is not fixed in the frequency-space, it moves
to higher frequency with time. Namely, the photon energy spectrum is
harder at later time.

\begin{figure}
\centering
\includegraphics[width=6cm]{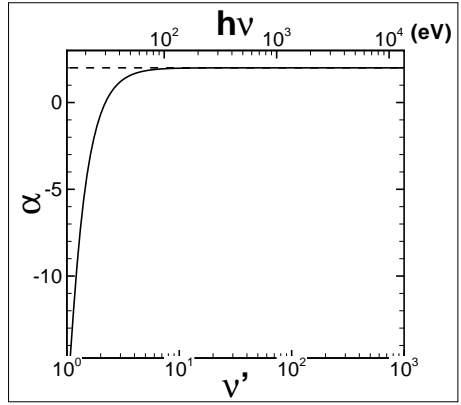}
\caption{$\alpha$ vs. $\log \nu/\nu_0$ for the source
$\dot{E}=5.8\times 10^{45}$ erg s$^{-1}$ at physical distance
$r=0.19$ (dash) and 0.24 (solid line) Mpc. Redshift is $1+z=10$.
The mesh in the numerical simulation is the same as that in Figure
\ref{fig3}.
}
\label{fig7}
\end{figure}

The spectral hardening can be measured by the index of power law
defined as
\begin{equation}
\label{eq31} \alpha=-{\partial \ln J \over \partial\ln \nu}.
\end{equation}
Figure \ref{fig7} plots $\alpha$ vs. $\nu$ for the frequency spectra of
Figure \ref{fig6}. Obviously, in the band $\nu/\nu_0<10$, $\alpha$
becomes smaller at larger $r$. Both Figures \ref{fig6} and \ref{fig7}
show that the frequency spectra of photons depend strongly on $t$
and $r$ when $t$ and $r$ are close to or inside the pre-heating layer.

\section{Discussions and conclusions}

We have described WENO scheme which is able to solve the
phase-space distribution function of photons in an isolated ionized
patch around individual source in the early stage of reionization.
This algorithm can produce robust results for the propagation of the
I- and T-fronts. It can also give stable results for the
time-dependent distribution of the small fraction of neutral
hydrogen within the I-front. We have developed the method to deal
with the stiffness of rate equations in time integration.
Consequently, the computational speed is acceptable. This algorithm
can be applied to problems with a wide range of intensity of
sources, from $10^{39}$ to 10$^{45}$ erg s$^{-1}$ with a power law
spectrum at redshift $1+z=10$. Since this algorithm treat the
frequency space as well as physical space, it can also be used for
sources with other frequency spectrum. We have not considered helium
and secondary effects in the atomic processes yet, however the
algorithm has no practical difficulty to include these factors.

We show that a common feature of the UV photon sources in the
reionization epoch is to form a preheating region. In the first stage
the I- and T-fronts in baryon gas are coincident, and propagate with
the speed about the same as the speed of light. When the frequency
spectrum of the UV photons is hardened, the evolution enters into the
second stage. The propagation speeds of both the ionizing and heating
fronts are less than that of light, but the T-front is always moving
faster than the ionizing front. In the spherical shell between the
I- and T-fronts, the kinetic temperature of gas can be as large
as $T\simeq 10^{3-4}$, while atoms are almost neural. Obviously,
the shell would be of interest in the search for the 21 cm
emission from the reionization epoch. The details of the preheating
region are sensitively dependent on the parameters of the problems.
For instance, the radius and thickness of the preheated shell do not
show a scaling relation on their dependence of the source intensity.
Therefore, an algorithm, which can properly handle models with
various parameters is necessary.

Using these results, one can make comments on some numerical
solvers used for the radiative transfer equations. Several solvers are
based on the approximation of omitting the time-derivative term of
the radiative transfer equation (Nakamoto et al. 2001; Cen 2002;
Razoumov et al. 2002; Rijkhorst et al. 2005; Susa 2006).  They
calculate the static ionization field for a given uniform or
non-uniform density fields of cosmic baryon matter. These codes
essentially are generalizing the static solution eq.(\ref{eq1}) to the
case of inhomogeneous background distribution of baryon matter, and
multiple sources. Such an approximation is useful to calculate the
ionization field for each given inhomogeneous mass field of baryon
matter, but it will no longer be a good approximation when the
retardation effect due to photon propagation is important. For
the problem of ionization profiles around a point source, the
retardation of photon is not always negligible. For instance, one
finds the following parameters have been used in the model: the ionization
and heating at comoving distance $r=10$ h$^{-1}$ Mpc from a point
source of age $t=0.5\times 10^6$ yrs at redshift $z=30$ (e.g. Cen,
2006). These parameters already violated the retardation constraint
$r\leq ct$.

The retardation effect can be seen from the time-dependence of
the I-front, $r_{\rm I}(t)$. If the retardation effect is negligible, an
analytical solution of $r_{\rm I}(t)$ is given by e.g. Whalen \&
Norman (2006)
\begin{equation}
\label{shuadd61}
r_{\rm I}(t)=R_s[1-e^{-t/t_{rec}}]^{1/3},
\end{equation}
where $R_s=(3\dot{N}/4\pi \alpha_{\rm HII} n_H^2)^{1/3}$ is the
static radius of the Str\"omgren sphere given by eq.(2), and
$t_{rec}=1/\alpha_{\rm HII} n_{\rm H}$ is recombination time scale.
According to eq.(\ref{shuadd61}), the time scale of the $r_{\rm
I}(t)$ evolution is $t_{rec}$, which is independent of the source
intensity $\dot{E}$. That is, $r_{\rm I}(t) \simeq R_s$ only if
$t\gg t_{rec}$, regardless $\dot{N}$. However, Figure \ref{fig4}
shows that time scale of the $r_{90}(t)$ evolution is
$\dot{N}$-dependent. This result also holds if we replace
$r_{90}(t)$ by, e.g. $r_{95}(t)$, and is true for a monochromatic
source too.

Considering the retardation effect, an analytical solution of
$r_{\rm I}(t)$ is approximately given by the following algebraic
equation (White et al. 2003)
\begin{equation}
\label{shuadd62}
r_{\rm I}(t)=\left [\frac{t-(r_{\rm I}(t)/c)}
   {4\pi n/3\dot{N}}\right ]^{1/3}.
\end{equation}
According to eq.(\ref{shuadd62}), when $t\ll t_c=(3\dot{N}/4\pi
nc^3)^{1/2}$, the speed of the I-front, $dr_{\rm I}(t)/dt \simeq c$,
and when $t\gg t_c$, $dr_{\rm I}(t)/dt \propto t^{-2/3}$. It is
qualitatively consistent with $dr_{90}(t)/dt$ shown in Figure
\ref{fig4}. First, the time scale of the $r_{90}(t)$ evolution is
approximately $\propto \dot{N}^{1/2}$, the larger the $\dot{N}$, the
longer the $t_c$. Second, $dr_{\rm I}(t)/dt$ decreases with $t$ by a
power law $dr_{90}(t)/dt\propto t^{-\beta}$ when $t$ is large.
However, the power law index $\beta > 2/3$. This is expected,
because eq.(\ref{shuadd62}) does not include the effect of
recombination, which leads to a slower decrease than
$dr_{90}(t)/dt$.

A common assumption used in eqs.(\ref{shuadd61}) and
(\ref{shuadd62}) is that the distribution $f_{\rm HI}(t,r)$ is
described by a step function like eq.(\ref{eq1}). Obviously, it
ignores the neutral hydrogen HI probably remained within $r_{\rm
I}(t)$. The tiny fraction of HI may also be hardly calculated well
by Monte-Carlo codes (Ciardi et al. 2001; Maselli et al. 2003),
which yield large numerical errors due to Possion shot noise.

The WENO algorithm has been successfully applied to kinetic
equations of the distribution function in the phase space with one
or two spatial dimensions and two or three phase space dimensions
with acceptable computational speed (Carrillo et al. 2006). We
believe that it is not difficult to implement the WENO algorithm for
radiative transfer problems with similar dimensions in the phase
space.

In our calculation, the evolution the cosmic baryon gas is not
tracked by the hydrodynamic equations, but is simply assumed to have a
uniform distribution with density $n$. This treatment would be
reasonable if the typical time scale of the relevant hydrodynamic
effects are less than that of the I- and T-fronts. For instance,
dynamical effects associated with sonic propagation would be
negligible in solving the propagation of the I- and T-fronts. Of
course, one can expect that richer results will be yielded if the
WENO scheme for radiative transfer problems can be incorporated with
the Euler hydrodynamics. Since the WENO scheme for the cosmological
hydrodynamical simulation has already been well established, it
would be possible to develop a unified
radiation/N-body/hydrodynamics code for cosmological problems.

\noindent{\bf Acknowledgments.}

This work is supported in part by the US NSF under the grants
AST-0506734 and AST-0507340. LLF acknowledges support from the
National Science Foundation of China under the grant 10573036.

\appendix

\section{Equations of Radiative Transfer}

For an ionized sphere associated with a point photon source, the
radiation transfer (RT) equation is (Bernstein, 1988, Qiu et al.
2006)
\begin{equation}
\label{eqa1}
        {\partial J\over\partial t} +\frac{1}{a}
        {\partial \over\partial x^i}(n^i J) -
        H\left(\nu{\partial J\over\partial\nu}-3J\right) =
        - k_\nu J + S.
\end{equation}
where $J(t, {\bf x}, \nu, n_i)$ is the specific intensity, $a$ the
cosmic factor, $H=\dot{a}/a$, $\nu$ the frequency of photon and
$n_i$ a unit vector in the direction of photon propagation. In
eq.(\ref{eqa1}), we take $c=1$. $k_\nu$ and $S$ are, respectively,
the absorption and sources of photons. The absorption coefficient
of eq.(\ref{eqa1}) is given by
\begin{equation}
k_{\nu} = \sigma(\nu)n_{{\rm HI}}(t, {\bf x})
\end{equation}
where the cross section $\sigma(\nu)=6.3\times 10^{-18}(\nu_0/\nu
)^3\ {\rm cm^2}$.

The number density of neutral hydrogen HI, $n_{\rm HI}(t,
{\bf x})$, is determined by
\begin{equation}
\label{eq22} \frac{df_{\rm HI}}{dt}=\alpha_{\rm HII}n_ef_{\rm HII}
-
  \Gamma_{\rm \gamma HI}f_{\rm HI}- \Gamma_{\rm e HI} n_ef_{\rm HI},
\end{equation}

Relevant parameters are taken from Theuns et al. (1998) as

1. recombination coefficient
\begin{equation}
\alpha_{\rm HII}=
 6.30\times10^{-11}T^{-1/2}T_3^{-0.2}/(1+T^{0.7}_6),
\end{equation}
where $T$ is temperature, and $T_n=T/10^n$.

2. collision ionization
\begin{equation}
\Gamma_{\rm eHI}=
    1.17\times 10^{-10}T^{1/2}e^{-157809.1/T}(1+T_5^{1/2})^{-1}
\end{equation}

3. photoionization
\begin{equation}
\Gamma_{\rm \gamma HI}(t, r)=
\int_{\nu_0}^{\infty} d\nu \frac{J(t,r, \nu)}{h\nu}\sigma(\nu).
\end{equation}

The temperature $T$ is determined by the equation
\begin{equation}
\label{eq14} n k_{\rm B}\frac{dT}{dt}= H-n^2 C.
\end{equation}

4. heating rate
\begin{equation}
\label{eq15} H = n_{HI}\int_{\nu_0}^{\infty}d\nu
   J(t,r,\nu)\sigma(\nu)\frac{\nu-\nu_0}{\nu}
\end{equation}
where $h\nu_0=2.176\times 10^{-11}$ ergs.

5. cooling. Since only the recombination
cooling is important, we have
\begin{eqnarray}
\label{eq16} C  & = & 8.70\times 10^{-27}T^{1/2}T_3^{-0.2}
   (1+T_6^{0.7})^{-1}[1-f_{\rm HI}]^2 \\ \nonumber
   & + & 1.42\times 10^{-27} T^{1/2}[1-f_{\rm HI}]^2 \\ \nonumber
   & + & 2.45\times 10^{-21}T^{1/2} e^{-157809.1/T}(1+T_5^{1/2})^{-1}
   (1-f_{\rm HI})f_{\rm HI}
     \\\nonumber
   & + & 7.5\times 10^{-19} e^{-118348/T}(1+T_5^{1/2})^{-1}
(1-f_{\rm HI})f_{\rm HI}
\end{eqnarray}
where $T_n=T/10^n$. The terms on the r.h.s. of eq.(\ref{eq16}) are,
respectively, the recombination cooling, collisional ionization
cooling, collisional excitation cooling and bremsstrahlung. Both
$H$ and $C$ are in the unit of ergs cm$^{3}$ s$^{-1}$.


\begin{thebibliography}{}

\bibitem{Alv} Alvarez, M., Bromm, V. \&  Shpiro, P. 2006, ApJ, 639, 621

\bibitem{Ber} Bernstein, J., 1988, Kinetic Theory in the Expanding
   Universe, Cambridge

\bibitem{C} Carrillo, J., Gamba, I., Majorana, A. \& Shu, C.-W.,
2006, J. Comp. Phys., 214, 55

\bibitem{Cen} Cen, R., 2002, ApJS, 141, 211

\bibitem{cenA} Cen, R., 2006, ApJ, 648, 47

\bibitem{CenH} Cen, R. \& Haiman, Z., 2000, ApJL, 542, L75

\bibitem{Chen} Chen, X \& Miralda-Escud\'e, J., 2006, astr-ph/0605439

\bibitem{Cia} Ciardi, B., Ferrara, A., Marri, S., \& Raimondo, G.
2001, MNRAS, 324, 381

\bibitem{Feng} Feng, L.L., Shu, C.-W., \& Zhang, M.P., 2004, ApJ, 612, 1

\bibitem{Gne} Gnedin, N.Y. \& Abel, T., 2001, NewA, 6, 437

\bibitem{IL} Iliev, I,T. et al. 2006, MNRAS, 371, 1057

\bibitem{Jia} Jiang, G. \& Shu, C.-W., 1996, J. Comp. Phys., 126, 202

\bibitem{KI} Kitayama, T., Yoshida, N., Susa, H. \& Umemura, M., 2004,
           ApJ, 613, 631

\bibitem{Jau} Madau, P. \& Rees, M., 2000, ApJL, 542, L69

\bibitem{Mase} Maselli, A., Ferrara, A. \& Ciardi, B. 2003, MNRAS,
345, 397

\bibitem{Mel} Mellema, G., Iliev, I.T., Alvarez, M.A. \& Shapiro, P.R.
2006, NewA, 11, 374

\bibitem{Naka} Nakamotoi, T., Umemura, M. \& Susa, H. 2001,
   MNRAS, 321, 593

\bibitem{Os} Osterbrock, D. \& Ferland, G. 2005, Astrophysics of
     Gaseous Nebulae and Active Galactic Nuclei, (University
     Science Books)

\bibitem{Qiu} Qiu, J.-M., Shu, C.-W., Feng, L.-L. \& Fang, L.Z.,
   2006, NewA, 12, 1

\bibitem{Raz} Razoumov, A., Norman M., Abel, T. \& Scott, D. 2002,
ApJ, 572, 695

\bibitem{Raz2} Razoumov, A. \& Scott, D. 1999, MNRAS, 309, 287

\bibitem{Ri} Ricotti, M., Gnedin, N., Shull, J. 2002, ApJ, 575, 33

\bibitem{Rij} Rijkhorst, E., Plewa, T., Dubey, A. \& Mellema, G. 2005,
astr-ph/0505213

\bibitem{sha} Shapiro, P.R., Iliev, I.T. \& Raga, A.C. 2004, MNRAS,
   348, 753

\bibitem{Shu2} Shu, C.-W., 2003, Int. J. Comp. Fluid Dyn., 17, 107

\bibitem{Shu} Shu, C.-W. \& Osher, S., 1988, J. Comp. Phys., 77, 439

\bibitem{sok} Sokasian, A. Abel, T. \& Hernquist, L.E. 2001, NewA,
6, 359

\bibitem{Str} Str\"omgren, B., 1939, ApJ, 89, 529

\bibitem{Susa} Susa, H. 2006, Pub. Astron. Soc. Japan. 2006, 58, 445

\bibitem{The} Theuns, T., Leonard, A., Efstathiou, G., Pearce, F.R. \&
   Thomas, P.A., 1998, MNRAS, 301, 478

\bibitem{To} Tozzi, P., Madau, P., Meiksin, A. \& Rees, M. 2000,
   ApJ, 528, 597.

\bibitem{wha1} Whalen, D. Abel, T., \& Norman, M. 2004, ApJ, 610, 14

\bibitem{wha} Whalen, D. \& Norman, M. 2006, ApJS, 162, 281

\bibitem{whit} White, R.L., Becker, R.H. \& Fan, X.H. 2003, ApJ, 126, 1

\bibitem{wyi1} Wyithe, J., Loeb, A., 2004, Nature, 427, 815

\bibitem{wyi} Wyithe, J. \& Loeb, A. \& Barnes, D., 2005, ApJ, 634,
715

\bibitem{Xu} Xu, Z. \& Shu, C.-W., 2005, J. Comp. Phys., 205, 458

\bibitem{yu} Yu, Q.J., 2005, ApJ, 623, 683

\bibitem{yulu} Yu, Q.J., \& Lu, Y.J., 2005, ApJ, 621, 31

\end{thebibliography}
\end{document}